\documentclass[prd,superscriptaddress,amsfonts,amssymb,amsmath,showpacs,twocolumn]{revtex4-2}
\usepackage{bm}
\usepackage{amsfonts}
\usepackage{latexsym}
\usepackage[latin1]{inputenc}
\usepackage{graphicx}
\usepackage{amsmath}
\usepackage{mathtools}
\usepackage{palatino}
\usepackage{mathpazo}
\usepackage{textcomp}
\usepackage{float}
\usepackage{booktabs}
\usepackage{dcolumn}
\usepackage{hyperref}
\usepackage{subfigure}
\usepackage[english]{babel}
\usepackage[autostyle]{csquotes}
\hypersetup{colorlinks,citecolor=red}
\usepackage{amsmath}
\usepackage{xcolor}
\usepackage{orcidlink}
\usepackage{commath}

\def\jnl@style{\it}
\def\aaref@jnl#1{{\jnl@style#1}}

\def\aaref@jnl#1{{\jnl@style#1}}

\def\aj{\aaref@jnl{AJ}}                   
\def\apj{\aaref@jnl{ApJ}}                 
\def\apjl{\aaref@jnl{ApJ}}                
\def\apjs{\aaref@jnl{ApJS}}               
\def\apss{\aaref@jnl{Ap\&SS}}             
\def\aap{\aaref@jnl{A\&A}}                
\def\aapr{\aaref@jnl{A\&A~Rev.}}          
\def\aaps{\aaref@jnl{A\&AS}}              
\def\mnras{\aaref@jnl{Mon.~Not.~Roy.~Astron.~Soc.}}             
\def\prd{\aaref@jnl{Phys.~Rev.~D}}        
\def\prc{\aaref@jnl{Phys.~Rev.~C}}  
\def\prl{\aaref@jnl{Phys.~Rev.~Lett.}}    
\def\qjras{\aaref@jnl{QJRAS}}             
\def\skytel{\aaref@jnl{S\&T}}             
\def\ssr{\aaref@jnl{Space~Sci.~Rev.}}     
\def\zap{\aaref@jnl{ZAp}}                 
\def\nat{\aaref@jnl{Nature}}              
\def\aplett{\aaref@jnl{Astrophys.~Lett.}} 
\def\apspr{\aaref@jnl{Astrophys.~Space~Phys.~Res.}} 
\def\physrep{\aaref@jnl{Phys.~Rep.}}      
\def\physscr{\aaref@jnl{Phys.~Scr}}       
\def\commat{\aaref@jnl{Comm.~Math.~Phys.}}              
\def\science{\aaref@jnl{Science}}               
\def\cqg{\aaref@jnl{Classical Quant.~Grav.}}            
\def\jpcs{\aaref@jnl{JPCS}}                                     
\def\ijmpd{\aaref@jnl{Int.~J.~Mod.~Phys.~D}}                    
\def\grg{\aaref@jnl{Gen.~Relat.~Gravit.}}               
\def\rpp{\aaref@jnl{Rep.~Prog.~Phys.}}          
\def\npa{\aaref@jnl{Nucl.~Phys.~A}}        
\def\lrr{\aaref@jnl{Living Rev.~Rel.}}                   
\def\jcap{\aaref@jnl{J.~Cosmology Astropart.~Phys.}}    
\def\rmp{\aaref@jnl{Rev.~Mod.~Phys.}}   
\def\epjc{\aaref@jnl{Eur.~Phys.~J.~C}}

\allowdisplaybreaks[1]

\begin{document}

\color{black}

\title{Gravastar model in Krori-Barua metric under $f(\mathcal{Q})$ gravity}

\author{Debasmita Mohanty\orcidlink{0009-0006-8118-5327}}
\email{newdebasmita@gmail.com}
\affiliation{Department of Mathematics, Birla Institute of Technology and
Science-Pilani,\\ Hyderabad Campus, Hyderabad-500078, India.}
\author{P.K. Sahoo\orcidlink{0000-0003-2130-8832}}
\email{pksahoo@hyderabad.bits-pilani.ac.in}
\affiliation{Department of Mathematics, Birla Institute of Technology and
Science-Pilani,\\ Hyderabad Campus, Hyderabad-500078, India.}

\date{\today}

\begin{abstract}
 In this paper, we explore the characteristics of a gravastar in $f(\mathcal{Q})$ gravity, which is upheld by Krori-Barua (KB) metric. We have used Krori-Barua (KB) metric for the interior and shell regions of the gravastar. We deduced our field equations by using Krori-Barua metric. In the outside regions of the gravastar, we have taken two regular black hole metrics. Additionally, we have applied the Israel junction condition to calculate the potential difference across the thin shell concerning different types of regular black holes, such as  Bardeen and Hayward. We have also discussed the physical properties like proper length, entropy, energy, EoS and stability.

\end{abstract}
\maketitle

\section{Introduction}\label{sec:I}
Gravastar is a promising alternative to conventional black holes. Unlike black holes, gravastars do not suffer from singularities, making them a safer and more reliable option. Additionally, the Modified Symmetric Teleparallel Equivalent of Gravity theory has shown remarkable success in studying the universe and celestial bodies, such as black holes and wormholes. These developments demonstrate the potential for significant breakthroughs in our understanding of the cosmos and the fundamental laws of nature.
The revolutionary concept of gravitational vacuum condensate stars, popularly known as gravastars, was first put forth by Mazur and Mottola \cite{Mottola/2002, Mazur/2004} in the early 21st century as a highly viable and advantageous substitute for conventional black holes. Black holes are notable entities in general relativity because they precisely solve Einstein's equations and exhibit entropy, a feature rooted in quantum mechanics. Roger Penrose's \cite{Penrose/1965} work established that collapses of a generic dast ball could still result in the emergence of singularities within black holes. Observational data from Ghez \cite{Ghez/1998} and Gillessen \cite{Gillessen/2009} further supported the existence of black holes or substantial compact objects. Their pioneering efforts earned Penrose, Ghez, and Gillessen the Nobel Prize in Physics in 2020. Nonetheless, the notion that a massive star's collapse terminally leads to a black hole is fraught with several issues, such as the exact nature of singularity and event horizon, which are still unknown.


The gravastar concept draws inspiration from phenomena observed in condensed matter physics, particularly during phase transitions, where a system's behaviour changes drastically, for example, shifting from ferromagnetic to paramagnetic states when approaching a critical temperature ($T\approx T_c$). This analogy extends to the behaviour of collapsing dust near the event horizon of a black hole, where quantum mechanical effects become significant, suggesting that such dust might act like a quantum many-body system. Furthermore, it's well-established that bosonic particles can form a Bose-Einstein condensate at extremely low temperatures. Leveraging this understanding, Mazur and Motalla proposed a novel explanation for dust collapse that circumvents the severe issues associated with traditional black hole models. The application of Bose-Einstein condensation in the astrophysical context has already been studied in the context of the Neutron star core and the formation of the Bose star. Several issues related to this have been highlighted in the studies by Panotopoulo \cite{Panotopoulo/2018}.\\
It has been observed that for an object of similar radius and mass as a black hole, the entropy can be represented as $S=\frac{4k+4}{7k+1}S_{BH}$, as highlighted in \cite{Zurek/1984}, where $k$ represents the equation of state (EoS) for the object. This particular form of expressing entropy is distinctive to gravastars, as it aligns with Hawking's semi-classical approach to black hole entropy on the surface level (where $k=1$), yet differs by being variable, unlike other non-singular astrophysical entities.\\
The concept of modified symmetric teleparallel gravity, also referred to as $f(\mathcal{Q})$ gravity, was introduced by Jimenez et al. \cite{Jimenez/2018}. This theory has been effectively applied in cosmology to account for the accelerated expansion of the universe in recent times \cite{Solanki/2021, Koussour/2022}. Additionally, numerous studies have analyzed observational data within the framework of $f(\mathcal{Q})$ gravity \cite{Mandal/2020,Gogoi/2023}. This gravitational theory has also significantly contributed to understanding the characteristics of various celestial phenomena, including black holes \cite{Ambrosio/2022} and wormholes \cite{Hassan/2023}, among others. It's worth mentioning that the investigation into gravastars within the context of alternative gravitational theories is a well-established area of research. Various works have explored gravastars in different gravitational theories, such as $f(R,T)$ gravity \cite{Das/2017, Pramit/2021}, $f(T)$ gravity \cite{Das/2020}, $f(\mathcal{Q})$ gravity \cite{Mohanty/2024}, and Rastall-Rainbow gravity \cite{Debnath/2021} to name a few. Ovgun et al., \cite{Ovgun/2017} studied the charged thin-shell gravastars in noncommutative geometry. One can also refer to \cite{Tayade/2022,Ghez/1998} for a detailed study of astrophysical objects. \\
In this article, we have developed the model under the $f(Q)$ gravity framework using the Krori-Barua (KB) metric. Numerous researchers have applied this approach to investigate various aspects of gravastars, wormholes, and strange stars within the frameworks of general relativity and other gravitational theories, as referenced in studies \cite{Biswas/2019, Bhar/2020, Sharif/2023}. The main characteristic of this solution is that it is completely free of singularities. A prominent aspect of this solution is its demonstration of a curvature singularity at its core, mirroring the singularity expected at a black hole's center, indicating its potential to probe black hole dynamics. However, it pertains to the vacuum surrounding the object. Furthermore, the KB solution enables the computation of gravitational redshift, which photons undergo as they depart from the object. This aspect is pivotal in astrophysics for assessing the mass and radius of dense celestial bodies like neutron stars and black holes. In essence, the KB solution emerges as a crucial instrument in astrophysics for delving into the actions of massive entities in space, providing a theoretical basis for examining black holes and comparable compact objects. It facilitates forecasts about their attributes, such as gravitational redshifts. The KB metric's value is attributed to its analytical approach, simplicity, compliance with energy conditions, versatility, and utility in comparative and validation studies. We also note that we only take the Krori-Barua metric as an interior region in the charged fluid sphere.\\
The interior solution of a charged ball, which is singularity-free, was first given by Krori and  Barua \cite{Krori/1975}. Even though an outside solution for such a situation is unique (By Brikhoff's theorem) and given by Ressner-Nordstrom solutions. The further properties of such solutions, like their relation with charge and mass and sufficient criteria for the singularity-free solution to exist, were studied by \cite{Junevicus/1976}. In this article, we have taken Ressner-Nordstrom's solution with Schwarzschild as well as regular black hole backgrounds like Bardeen and Hayward's black hole backgrounds. For the interior, we have taken the Krori-Barua metric to mimic the original motivation for the metric. So, with the outside Ressner-Norstrom solution and Krori-Barua metric in the interior and shell, we have given a full understanding of the charged gravastar's physical properties like proper length, entropy, energy, EoS parameter and stability, etc. Also, we have given phenomenological predictions about the potential across this shell, which can later be verified via the shadow of the gravastar via future radio telescopes.\\
In this study, we focus on two varieties of regular black holes to form the shell's outer layer. This choice is driven by the hypothesis that the singularity issue of black holes can be addressed by adjusting specific metric coefficients. The main motivation for taking regular black hole outside layer is that the inspiration behind the Bardeen black hole \cite{Bardeen/1968} stems from the possibility of a singularity-free metric that fulfils both the conditions of asymptotic flatness and the weak energy condition, while also having a regular center. We've also analyzed the situation when the exterior is given by the Hayward black hole \cite{Hayward/2006}. Hayward metric was inspired by how masses around the Bardeen black hole might accumulate via a ``Vaidya'' like solution.\\
Gravastar consists of three parts: the core, a thin layer in the middle, and the surrounding space. Each part has its own set of boundaries, denoted by inner $(r_1)$ and outer $(r_2)$ radii, where $r_1<r_2$. The composition of these layers is as follows:
\begin{itemize}
\item The core region extends from the center up to $r_1$ and behaves according to a specific rule, where pressure is the negative of density, i.e.$(p=-\rho)$.
\item  The middle thin layer stretches from $r_1$ to $r_2$, acting as a bridge made of a very rigid fluid, where pressure equals density $(p=\rho)$.
\item  The outer space starts from $r_2$ onwards, with no pressure, resembling the empty universe described by Schwarzschild's equations.
\end{itemize}

The essence of the present study can be summarized as: In section \ref{sec:II}, we present the construction of $f(\mathcal{Q})$ gravity. In section \ref{sec:III}, we discuss Maxwell's equation in curved space-time. In section \ref{sec:IV}, we utilize the field equation equations to analyze a spherically symmetric spacetime, and by choosing the Krori-Barua metric potential, we deduced our field equation. In section \ref{sec:V}, we discuss the structure of charged gravastar. In section \ref{sec:VI}, we address the boundary condition. In section \ref{sec:VII}, we looked at the junction condition.In section \ref{sec:VIII}, we discussed the physical features of our model. In section \ref{sec:IX}, we provide the conclusion of our analysis.

\section{Construction of $f(Q)$ Gravity} \label{sec:II}

In $f(\mathcal{Q})$ gravity theory, the metric tensor $g_{\mu \nu}$ and connection $\Gamma^\lambda_{\mu \nu}$ are considered as independent variables in a metric-affine space-time. The non-metricity of the connection in this theory is determined by,

\begin{equation}\label{eq:1} 
    \mathcal{Q}_{\alpha \mu \nu}=\bigtriangledown_{\alpha} g_{\mu \nu}=\partial_{\alpha} g_{\mu \nu} -\Gamma^{\lambda}\,_{\alpha \mu} \,g_{\lambda \nu}- \Gamma^{\lambda}\,_{\alpha \nu} g_{\mu \lambda} .
\end{equation}

Thus, in essence, the affine relationship can be made up of the three separate parts given below:

\begin{equation}\label{eq:2}
\Gamma^{\lambda}\,_{\mu \nu}=  \{ {}^\lambda\,{}_{\mu \nu}\}+ K^{\lambda}\,_{\mu \nu} +L^{\lambda}\, _{\mu \nu}. 
\end{equation}

The Levi-Civita connection and contortion are represented by the symbols $\{ {}^\lambda\,{}_{\mu \nu}\}$  and $K^{\lambda}\,_{\mu \nu}$, respectively, and are defined as follows:

\begin{equation} \label{eq:3}
    \{ {}^\lambda\,{}_{\mu \nu}\} \equiv \frac{1}{2} g^{\lambda \beta} \left( \partial_{\mu} g_{\beta \nu}+\partial_{\nu} g_{\beta \mu}-\partial_{\beta} g_{\mu \nu} \right),
\end{equation}

\begin{equation}\label{eq:4}
K^{\lambda}\,_{\mu \nu} \equiv \frac{1}{2} T^{\lambda}\,_{\mu \nu}+ T_(\mu^{\lambda}\,_\nu) ,
\end{equation}  

Where the antisymmetric portion of the affine connection is denoted by $T^{\lambda}\,_{\mu \nu}$ and is defined as follows: $T^{\lambda}\,_{\mu \nu} \equiv 2 \Gamma^{\lambda}\,_{[\mu \nu]}$.
$L^{\lambda}\,_{\mu \nu}$ is the disformation described by

\begin{equation}\label{eq:5}
 L^{\lambda}\,_{\mu \nu} \equiv \frac{1}{2} \mathcal{Q}^{\lambda}\,_{\mu \nu}- \mathcal{Q}(_{\mu}\,^{\lambda}\,_{\nu} ) .
\end{equation}

Next, we define the non-metricity conjugate as follows:

\begin{multline}\label{eq:6}
    P^{\alpha}\,_{\mu \nu}=-\frac{1}{4} \mathcal{Q}^{\alpha}\,_{\mu \nu}+\frac{1}{2} \mathcal{Q}(_{\mu}\,^{\alpha}\,_{\nu})+\frac{1}{4}(\mathcal{Q}^{\alpha}-\Tilde{\mathcal{Q}}^{\alpha}) g_{\mu \nu} \\ 
    -\frac{1}{4} \delta^{\alpha}(_\mu \mathcal{Q}_{\nu}), 
\end{multline}

where the two independent traces of the non-metricity tensor are $\mathcal{Q}_{\alpha} \equiv \mathcal{Q}_{\alpha}\,^{\mu}\,_{\mu}$ and  $\Tilde{\mathcal{Q}}_{\alpha} \equiv \mathcal{Q}^{\mu}\,_{\alpha \mu}$

Lastly, we define the non-metricity scalar as follows:

\begin{equation}\label{eq:7}
    \mathcal{Q}\equiv - \mathcal{Q}_{\alpha \mu \nu} P^{\alpha \mu \nu} .
\end{equation}

$f(\mathcal{Q})$ gravity is determined by the action \cite{Jimenez/2018} along Lagrange multipliers.

\begin{multline}\label{eq:8}
    S=\int \sqrt{-g} d^4x \left[\ \frac{1}{2} f(\mathcal{Q}) + \lambda_{\alpha}\,^{\beta \mu \nu} R^{\alpha}\,_{\beta \mu \nu} + \lambda_{\alpha}\,^{\mu \nu} T^{\alpha}\,_{\mu \nu}\right.\\
    \left.+ \mathcal{L}_m + \mathcal{L}_e \right]\, .
\end{multline}

The determinant of the metric $g_{\mu \nu}$ is denoted by $g$, while the arbitrary function of the non-metricity $\mathcal{Q}$ is represented by $f(\mathcal{Q})$. The Lagrange multipliers, matter Lagrangian, and Lagrangian for the electromagnetic are represented by $\lambda_{\alpha}\,^{\beta \mu \nu}, \mathcal{L}_m, $ and $\mathcal{L}_e$, respectively.

By changing action \eqref{eq:8} to the metric \cite{Wang/2022}, the field equation may be obtained.

\begin{multline}\label{eq:9}
-T_{\mu \nu} + E_{\mu \nu}=\frac{2}{\sqrt{-g}} \bigtriangledown_{\alpha} \left( \sqrt{-g} f_{\mathcal{Q}} P^{\alpha}\,_{\mu \nu}\right) +\frac{1}{2} g_{\mu \nu} f\\
+f_{\mathcal{Q}} \left(  P_{\mu \alpha \beta} {\mathcal{Q}}_{\nu} \,^{\alpha \beta}  - {2\mathcal{Q}}_{\alpha \beta \mu} P^{\alpha \beta}\,_{\nu}\right), 
\end{multline}

where $f_{\mathcal{Q}} \equiv \partial_{\mathcal{Q}} f(\mathcal{Q})$ is the derivative of $f(\mathcal{Q})$ with regard to $\mathcal{Q}$, and $E_{\mu \nu}$ is the energy-momentum tensor of the electromagnetic field.The following defines the energy-momentum tensor:

\begin{equation}\label{eq:10}
   T_{\mu \nu} \equiv - \frac{2}{\sqrt{-g}} \frac{\delta(\sqrt{-g} \mathcal{L}_m)}{\delta g ^{\mu \nu}}.
\end{equation}

It is observed that the energy-momentum tensor of the electromagnetic field, $E_{\mu \nu}$, may be located as \cite{Yousaf/2019}.

\begin{equation} \label{eq:11}
    E_{\mu \nu} (EM)= \frac{1}{4 \pi} \left[- F^\zeta_\mu F_{\nu \zeta} +\frac{1}{4} F_{\eta\zeta} F^{\eta\zeta} g_{\mu \nu}\right] .
\end{equation}

Now, by changing equation \eqref{eq:8} in relation to the connection we find,

\begin{equation}\label{eq:12}
\bigtriangledown _{\rho} \lambda _{\alpha}\,^{\nu \mu \rho} + \lambda_{\alpha} \, ^{\mu \nu} = \sqrt{-g} f_{\mathcal{Q}} P^{\alpha}\,_{\mu \nu} +H_{\alpha}\, ^{\mu \nu}, 
\end{equation}
where the hyper momentum tensor density, $H_{\alpha}\, ^{\mu \nu}$, is defined as follows \cite{Wang/2022}:

\begin{equation} \label{eq:13}
   H_{\alpha}\, ^{\mu \nu} = \frac{-1}{2} \frac{\delta (\mathcal{L}_m + \mathcal{L}_e)}{\delta \Gamma^{\alpha}\,_{\mu \nu}}. 
\end{equation}

Equation \eqref{eq:12} can be reduced using the anti-symmetry property of $\mu$ and $\nu$ in the Lagrangian multiplier coefficients.

\begin{equation}\label{eq:14}
    \bigtriangledown_{\mu} \bigtriangledown_{\nu} \left( \sqrt{-g} f_{\mathcal{Q}} P^{\mu \nu}\,_{\alpha} +H_{\alpha}\, ^{\mu \nu}\right) =0. 
\end{equation}

Assuming $\bigtriangledown_{\mu} \bigtriangledown_{\nu} H_{\alpha}\, ^{\mu \nu}=0$. The above equation becomes \cite{Jimenez/2018}., 

\begin{equation} \label{eq:15}
    \bigtriangledown_{\mu} \bigtriangledown_{\nu} \left( \sqrt{-g}f_{\mathcal{Q}} P^{\mu \nu}\,_{\alpha} \right)=0. 
\end{equation}
The affine connection takes on the following form when there is no curvature or torsion.

\begin{equation} \label{eq:16}
    \Gamma^{\alpha}\,_{\mu \nu} = \left( \frac{\partial x^{\alpha}}{\partial \theta^{\lambda}} \right) \partial_{\mu} \partial_{\nu}\, \theta^{\lambda}. 
\end{equation}

In this instance, $\Gamma^{\alpha}\,_{\mu \nu}=0$ can be achieved by selecting a unique set of coordinates, known as the coincident gauge. Non-metricity so reduces to,

\begin{equation}\label{eq:17}
    {\mathcal{Q}}_{\alpha \mu \nu} = \partial_{\alpha} g_{\mu \nu}. 
\end{equation}

As a result, the computation is much simplified because only the metric variable is necessary. With the exception of STGR \cite{Beltran/2020}, the action is no longer diffeomorphism invariant. The covariant formulation of $f(\mathcal{Q})$ gravity can be used to get around the problem. As the affine connection in equation \eqref{eq:16} is only inertial, one way to use the covariant formulation is to first find the affine connection without gravity \cite{Zhao/2022}. But as this study shows, the coincident gauge's off-diagonal field equations component would place severe restrictions on $f(\mathcal{Q})$ gravity, yielding nontrivial functional forms for $f(\mathcal{Q})$.

\section{Maxwell's equation in curved space time} \label{sec:III}

The four Maxwell equations can all be written as, 

\begin{equation} \label{eq:18}
    \frac{\partial F_{\mu \nu}}{\partial x_{\nu}}=\mu_0 J_{\mu}.
\end{equation}

\begin{equation} \label{eq:19}
    \frac{\partial F_{\mu \nu}}{\partial x_{\lambda}} +\frac{\partial F_{\nu \lambda}}{\partial x_{\mu}} +\frac{\partial F_{\lambda \mu }}{\partial x_{\nu}} =0.
\end{equation}

Lagrangian density is defined as follows for a free electromagnetic field:
\begin{equation} \label{eq:20}
  L_{EM} =-\frac{1}{16 \pi} F^{\mu \nu} F_{\mu \nu},
\end{equation}
with
\begin{equation} \label{eq:21}
  F_{\mu \nu}= A_{\nu,\mu}  - A_{\mu , \nu}.
\end{equation}

We observe that the Euler-Lagrange equation yields Maxwell's equation if we modify the Lagrangian to $A_{\mu}$.\\

Thus, we can obtain the energy-momentum tensor of the aforementioned Lagrangian by changing it with respect to the metric,

\begin{equation} \label{eq:22}
  E_{\mu \nu}=2 \frac{\partial \mathcal{L}_m}{\partial g^{\mu \nu}} - \mathcal{L}_m g_{\mu \nu}.
\end{equation}
Inserting the EM Lagrangian density, we so obtain,
\begin{equation*} 
    E_{\mu \nu} (EM)= \frac{1}{4 \pi} \left[- F^{\zeta}_\mu F_{\nu \zeta} +\frac{1}{4} F_{\eta \zeta} F^{\eta \zeta} g_{\mu \nu}\right] 
\end{equation*}

We observe that $J^\mu$ is zero for an isolated static charge. We also observe that because we are using spherical polar coordinates, the radial potential is the only component of interest for the field strength of a static charge.

\begin{equation*}
    A_0=\phi \neq 0 , \,\, A_2=A_3=A_1=0.
\end{equation*}

Thus, only $F_{01}, F_{10}$ will be nonzero among all the components of $F_{\mu \nu}$. \\
Thus, the electromagnetic field strength tensor can be written as follows:

\begin{equation*} 
    F_{\mu \nu}=\phi_{,1}  \begin{pmatrix}
    0 & -1 & 0 & 0 \\
    1 & 0 & 0 & 0 \\
    0 & 0 & 0 & 0 \\
    0 & 0 & 0 & 0
\end{pmatrix}
\end{equation*}

Using Maxwell's equation, we obtain:

\begin{equation} \label{eq:23}
\begin{gathered}
    \frac{\partial F_{\mu \nu}}{\partial x_{\nu}}= {\mu}_0 J_{\mu}=0 , \\
    \implies (g^{00} g^{11} F_{01} \sqrt{-g})_{,1} =0, \\
    \implies  (g^{00} g^{11} \phi \sqrt{-g})_{,1} =0,\\
    \implies e^{-\frac{\lambda +\nu}{2}} \phi_{,1} r^2 = Q = constant.
\end{gathered}
\end{equation}
Consequently, the electromagnetic field's energy-momentum tensor is expressed as follows:
\begin{equation} \label{eq:24}
    E^{\mu}_{\nu} = \frac{1}{8 \pi} diag (\frac{Q^2}{r^4}) (1,1,-1,-1).
\end{equation}
It should be noted that to obtain the graph, we utilize $Q$ as a dimensionless parameter in this case. Nevertheless, $\frac{Q^2G}{4\pi \epsilon_0c^4}$ provides the dimension expression for $Q$.

\section{$f(\mathcal{Q})$ gravity with spherical symmetry and choice of the metric potential} \label{sec:IV}

We will now analyze the spherically symmetric configurations. In spherical coordinates, the metric adheres to the conventional format. 

\begin{equation}\label{eq:25}
 ds^2=  - e^{A(r)} dt^2 + e^{B(r)} dr^2 + r^2 (d \theta^2 + \sin^2 \theta d\phi^2).
\end{equation}

Moreover, the non-metricity scalar $\mathcal{Q}$  is

\begin{equation}\label{eq:26}
    \mathcal{Q}(r)=-\frac{2 e^{-B}}{r} \left(  A' + \frac{1}{r}\right) . 
\end{equation}
In $f(\mathcal{Q})$ gravity, the Einstein-Maxwell field equation is provided by,

\begin{widetext}
\begin{equation}\label{eq:27}
    \frac{f}{2}+ \frac{2}{r} e^{-B} f_{\mathcal{Q} \mathcal{Q}} {\mathcal{Q}}' -f_{\mathcal{Q}}\biggl[\mathcal{Q}+\frac{e^{-B}}{r} (A'+B')+\frac{1}{r^2}\biggr] = \rho + 2\pi E^2, 
\end{equation}

\begin{equation}\label{eq:28}
    -\frac{f}{2}+f_{\mathcal{Q}}\left(  \mathcal{Q}+\frac{1}{r^2}\right)= p_r - 2 \pi E^2, 
\end{equation}
\begin{equation}\label{eq:29}
    -\frac{f}{2}- e^{-B} \left(  \frac{1}{r} +\frac{A'}{2}\right) f_{\mathcal{Q} \mathcal{Q}} {\mathcal{Q}}' + f_{\mathcal{Q}}\biggl[\frac{\mathcal{Q}}{2}-e^{-B} \left({\frac{A''}{2}+\left( \frac{A'}{4} +\frac{1}{2r}\right) \left( A'-B'\right)} \right)\biggr] =p_t + 2 \pi E^2, 
\end{equation}
and
\begin{equation}\label{eq:30}
  \frac{\cot{\theta}}{2}  f_{\mathcal{Q} \mathcal{Q}} {\mathcal{Q}}'=0. 
\end{equation}
\end{widetext}
\begin{equation}\label{eq:31}
\left( r^2 E\right)'= 4 \pi r^2 \sigma(r) e^{\frac{B}{2}}. 
\end{equation}

Therefore, the invariant energy is,
\begin{equation}\label{eq:32}
    E(r)=\frac{1}{r^2} \int{4 \pi r^2  \sigma(r) e^{\frac{B}{2}} dr}. 
\end{equation}

The matter density, isotropic pressure, and electric field of a charged fluid sphere are indicated by the terms $\rho$, $p$, and $E(r)$, respectively. Additionally, the thin shell's surface energy density is shown by $\sigma(r)$. Differentiation to the radial coordinate, $r$, is shown by the prime notation.\\
We select the coefficients of $g_{rr}$ and $g_{tt}$ for our current model as,
\begin{equation}\label{eq:33}
    e^{B(r)}=e^{Xr^2},\,\,\,\, e^{A(r)}=e^{Yr^2+Z},
\end{equation}
where $X, Y$ and $Z$ are unknown constants whose value can be obtained using matching conditions. And they fulfil the Einstein-Maxwell equations, representing a charged spherical body composed of a perfect fluid. 

The field equation \eqref{eq:30} requires off-diagonal terms to be satisfied. Therefore, we can only consider $f(\mathcal{Q})$ in the following form:

\begin{equation} \label{eq:34}
    f(\mathcal{Q})= a{\mathcal{Q}}+b ,
\end{equation}
with $a$ and $b$ being constants.  We note that at $a=-1$ and $b=0$, it reduces to Einstein's GR. 

Now using \eqref{eq:33} and \eqref{eq:34} in equations \eqref{eq:27}-\eqref{eq:29} we obtained the following equations.
\begin{equation} \label{eq:35}
    \frac{b}{2}-\frac{a}{r^2}-e^{-Xr^2}\left[a\left( 2X-\frac{1}{r^2}\right)\right]=\rho+2 \pi E^2 ,
\end{equation}

\begin{equation}\label{eq:36}
   - \frac{b}{2}+\frac{a}{r^2}-e^{-Xr^2} \left[a \left(2Y+\frac{1}{r^2}\right)\right] =p_r -2\pi E^2 ,
\end{equation}

\begin{equation} \label{eq:37}
    -\frac{b}{2}+e^{-Xr^2}\left[a\left( X-2Y+Y(X-Y)r^2  \right)   \right]=p_t+ 2\pi E^2.
\end{equation}
By solving the above equations, we obtained
\begin{equation} \label{eq:38}
    p=-\frac{b}{2}-\frac{a}{2r^2}+a e^{-Xr^2} \left[ 2Y+\frac{1}{2r^2}-\frac{X}{2}-\frac{Y}{2} (X-Y)r^2\right],
\end{equation}

\begin{equation} \label{eq:39}
    \rho = \frac{b}{2} +\frac{a}{2r^2}+ a e^{-Xr^2} \left[ \frac{5X}{2}-\frac{1}{2r^2}+\frac{Y}{2} (X-Y)r^2\right],
\end{equation}

\begin{equation} \label{eq:40}
    E^2=\frac{1}{\pi} \left[  \frac{a}{4r^2}-\frac{a}{4} e^{-Xr^2} \left( \frac{1}{r^2}+X+Y(X-Y)r^2\right)\right].
\end{equation}

\section{Structure of charged gravastar} \label{sec:V}

In this section, we examine three different regions of the gravastar in the presence of an electromagnetic field.

\begin{enumerate}
    \item [(A).] Interior region $\implies p=- \rho$ ,
    \item[(B).] Thin shell $\implies p=\rho $ ,
    \item[(C).] Exterior region $\implies p=0$.
\end{enumerate}

\subsection{Interior region}

The basic cosmic EoS $p = \omega \rho$, where $\omega$ is the EoS parameter that takes a variable value for different regions, is followed by the three distinct zones in the fundamental model provided by Mazur and Motola \cite{Mottola/2002, Mazur/2004}. Here, we assume the presence of an intriguing gravitational source in the interior region. Despite the potential that they are both only different representations of the same thing, dark matter and dark energy are typically thought to be distinct phenomena. We are interested in examining the EoS to characterize the dark sector in the interior region, which is provided by
\begin{equation} \label{eq:41}
    p=-\rho .
\end{equation}

Using (\ref{eq:41}) in equations (\ref{eq:38}) and (\ref{eq:39}) we obtain

\begin{equation} \label{eq:42}
    ae^{-Xr^2} (Y+X)=0 .
\end{equation}

Again, by using the above equation, we obtain an expression for the charge as  

\begin{equation} \label{eq:43}
    E^2= \frac{1}{\pi} \left[\frac{a}{4r^2} -\frac{a}{4}  e^{-X r^2}  \left(\frac{1}{r^2}+X-2X^2r^2 \right)      \right].
\end{equation}

The active gravitational mass $M(r)$ can be obtained as the following formula,
\begin{widetext}
\begin{multline}\label{eq:44}
    M(r)=\frac{1}{12}\pi\left[8 b r^3+\frac{9 \sqrt{\pi } a \left(2 X^2+X Y-Y^2\right) \text{erf}\left(\sqrt{X} r\right)}{X^{5/2}} - \frac{1}{X^2} \left(6 a r e^{-X r^2} \left(X^2 \left(-8 e^{X r^2}+2 Y r^2+14\right)+4 X^3 r^2 -3 Y^2 \right. \right. \right.\\ \left.\left.\left. +X Y \left(3-2 Y r^2\right)\right) \right)\right].
\end{multline}
\end{widetext}

\subsection{Shell}

Here, we consider a stiff perfect fluid that satisfies the EoS 
\begin{equation}\label{eq:45}
p =\rho .
\end{equation}
to be contained in the thin shell. 
This EoS is a special case of a barotropic EoS with $\omega = 1$ and $p = \omega\rho$. Barotropic fluids are ones in which $p = p(\rho)$, or the pressure and density alone rely on each other. Although they are seen as improbable, their simplicity provides the pedagogical benefit of demonstrating the variety of approaches used to address different systems and "physically" fascinating issues. In this context, we note that Zel'dovich \cite{Zeldovich/1972} initially introduced the idea of this kind of fluid in connection with a cold baryonic universe, describing it as a stiff fluid.
Staelens et al. \cite{Staelens/2021} investigated the spherical collapse of an over-density of a barotropic fluid with a linear equation of state in a cosmic backdrop.  Several astrophysics and cosmology researchers have already employed the stiff fluid model \cite{Carr/1975, Braje/2002, Madsen/1992, Ferrari/2007,Rahaman/2014}. It can be observed that the field equations in the non-vacuum region or the shell are quite difficult to solve. Nonetheless, an analytical solution was obtained within the thin shell limit's parameters, i.e., $0<e^{-B(r)}<<1$. We can contend that, as proposed by Israel \cite{Israel}, the inner region between the two space-times must consist of a thin shell. Furthermore, any parameter that is a function of $r$ can be regarded as $<< 1$ as $r\to 0$ in general. The approximation of this kind results in the following reduction of our field: Eq.(\ref{eq:27})-Eq. (\ref{eq:29})  to:

\begin{equation}\label{eq:46}
    \frac{f}{2}-f_Q\left[Q+\frac{e^{-B}}{r}B'+\frac{1}{r^2}\right]=\rho +2\pi E^2,
\end{equation}

\begin{equation}\label{eq:47}
     -\frac{f}{2}+f_{\mathcal{Q}}\left(  \mathcal{Q}+\frac{1}{r^2}\right)= p_r - 2 \pi E^2, 
\end{equation}

\begin{equation}\label{eq:48}
    -\frac{f}{2}+f_Q \left[\frac{Q}{2}-e^{-B}\left(\frac{-A'B'}{4}-\frac{B'}{2r}\right)   \right]=p_t+2\pi E^2.
\end{equation}

By utilizing equations \eqref{eq:46} and \eqref{eq:47} in equations \eqref{eq:45} through \eqref{eq:38} and \eqref{eq:39}, we obtain the metric potential.
\begin{equation} \label{eq:49}
    e^{-B(r)}=-e^{-Xr^2}-c_1,
\end{equation}
where $c_1$ is integration constant.

\subsection{Exterior region}
\subsubsection{ Reissner-Nordstrom (R-N) metric}
It is assumed that the exterior of the charged gravastar obeys the EoS $p=0$, demonstrating complete vacuum sealing of the shell. The external region is described by the Reissner-Nordstrom (R-N) line element given by

\begin{multline} \label{eq:50}
ds^2=-\left( 1-\frac{2M}{r}+\frac{Q^2}{r^2}\right) dt^2 +\frac{dr^2}{ \left(1-\frac{2M}{r} +\frac{Q^2}{r^2}\right)} +r^2(d \theta^2  \\
+ \sin^2 \theta d\phi^2). 
\end{multline}
Where $M$ and $Q$ represent the mass and charge of the gravatar.
\subsubsection{Regular black holes}
As for the exterior geometry, we select two different regular BHs.
\begin{equation}\label{eq:51}
    ds^2= -F(r) dt^2 + F(r)^{-1} dr^2 + r^2 d{\theta}^2 + r^2 \sin^2 \theta d{\phi}^2.
\end{equation}

\begin{itemize}
\item $F(r)$ represents Bardeen black holes if, $F(r)= 1-\frac{2Mr^2}{(r^2+e^2)^{\frac{3}{2}}} +\frac{Q^2}{r^2}$.
    
    \item $F(r)$ represent Hayward black holes if, $F(r)=1-\frac{2Mr^2}{r^3+2Ml^2}+\frac{Q^2}{r^2}$.

\end{itemize}

\section{Boundary condition}\label{sec:VI}

In this part, we match our internal spacetime to the external spacetime to fix the constants $X, Y,$ and $Z$.

A series of relations is obtained between the inner and exterior regions of the boundary surface $r = R$, based on the continuity of the metric coefficients $g_{tt}, g_{rr}$, and $\frac{\partial g_{tt}}{\partial r}$.

\subsection{Reissner-Nordstrom metric}
\begin{equation} \label{eq:52}
    1-\frac{2M}{R}+\frac{Q^2}{R^2}=e^{YR^2+Z},
\end{equation}
\begin{equation} \label{eq:53}
    1-\frac{2M}{R}+\frac{Q^2}{R^2}=e^{-XR^2},
\end{equation}

\begin{equation} \label{eq:54} 
    \frac{M}{R^2} - \frac{Q^2}{R^3}=YR e^{YR^2+Z}.
\end{equation}
The values of constants $X$, $Y$, and $Z$ in terms of total mass $M$, radius $R$, and charge $Q$ are found using equations \eqref{eq:52}-\eqref{eq:54}. When the aforementioned set of equations is solved, we obtain
\begin{equation} \label{eq:55}
    X=-\frac{1}{R^2} \ln \left(1-\frac{2M}{R}+\frac{Q^2}{R^2}\right),
\end{equation}
\begin{equation} \label{eq:56}
    Y= \frac{1}{R^2}\left(1-\frac{2M}{R}+\frac{Q^2}{R^2}\right)^{-1}\left(\frac{M}{R}-\frac{Q^2}{R^2} \right),
\end{equation}

\begin{equation}\label{eq:57}
    Z=\ln \left(   1-\frac{2M}{R}+\frac{Q^2}{R^2}\right)-\frac{\frac{M}{R}-\frac{Q^2}{R^2}}{1-\frac{2M}{R}+\frac{Q^2}{R^2}}.
\end{equation}

\subsection{Bardeen black holes}

\begin{equation} \label{eq:58}
    1-\frac{2MR^2}{(R^2+e^2)^{\frac{3}{2}}}+\frac{Q^2}{R^2}= e^{YR^2+Z},
\end{equation}
\begin{equation} \label{eq:59}
  1-\frac{2MR^2}{(R^2+e^2)^{\frac{3}{2}}}+\frac{Q^2}{R^2}=e^{-XR^2},  
\end{equation}
\begin{equation} \label{eq:60}
   \frac{6 M R^3}{\left(e^2+R^2\right)^{5/2}}-\frac{4 M R}{\left(e^2+R^2\right)^{3/2}}-\frac{2 Q^2}{R^3}=YR e^{YR^2+Z}.
\end{equation}
The result of solving the aforementioned system of equations is

\begin{equation} \label{eq:61}
    X=\frac{-1}{R^2} \ln{\left(1-\frac{2MR^2}{(R^2+e^2)^{\frac{3}{2}}}+\frac{Q^2}{R^2}\right)},
\end{equation}

\begin{equation} \label{eq:62}
    Y=-\frac{R \left(\frac{2 Q^2}{R^3}-\frac{2 M \left(R^3-2 e^2 R\right)}{\left(e^2+R^2\right)^{5/2}}\right)}{-\frac{2 M R^4}{\left(e^2+R^2\right)^{3/2}}+Q^2+R^2},
\end{equation}

\begin{multline} \label{eq:63}
    Z=\frac{R^3 \left(\frac{2 Q^2}{R^3}-\frac{2 M \left(R^3-2 e^2 R\right)}{\left(e^2+R^2\right)^{5/2}}\right)}{-\frac{2 M R^4}{\left(e^2+R^2\right)^{3/2}}+Q^2+R^2}+\log \left(-\frac{2 M R^2}{\left(e^2+R^2\right)^{3/2}} \right. \\ \left.
    +\frac{Q^2}{R^2}+1\right).
\end{multline}

\subsection{Hayward black holes}
\begin{equation} \label{eq:64}
    1-\frac{2MR^2}{R^3+2Ml^2}+\frac{Q^2}{R^2}= e^{YR^2+Z},
\end{equation}

\begin{equation} \label{eq:65}
    1-\frac{2MR^2}{R^3+2Ml^2}+\frac{Q^2}{R^2}=e^{-XR^2}, 
\end{equation}

\begin{equation}\label{eq:66}
    -\frac{4 M R}{2 l^2 M+R^3}+\frac{6 M R^4}{\left(2 l^2 M+R^3\right)^2}-\frac{2 Q^2}{R^3}=YR e^{YR^2+Z}.
\end{equation}
The result of solving the aforementioned system of equations is
\begin{equation} \label{eq:67}
    X=-\frac{1}{R^2} \ln{\left(  1-\frac{2MR^2}{R^3+2Ml^2}+\frac{Q^2}{R^2}   \right)},
\end{equation}

\begin{equation}\label{eq:68}
    Y=-\frac{\frac{2 M R \left(R^3-4 l^2 M\right)}{\left(2 l^2 M+R^3\right)^2}-\frac{2 Q^2}{R^3}}{R \left(\frac{2 M R^2}{2 l^2 M+R^3}-\frac{Q^2}{R^2}-1\right)},
\end{equation}

\begin{equation}\label{eq:69}
    Z=\frac{R \left(\frac{2 M R \left(R^3-4 l^2 M\right)}{\left(2 l^2 M+R^3\right)^2}-\frac{2 Q^2}{R^3}\right)}{\frac{2 M R^2}{2 l^2 M+R^3}-\frac{Q^2}{R^2}-1}+\log \left(-\frac{2 M R^2}{2 l^2 M+R^3}+\frac{Q^2}{R^2}+1\right).
\end{equation}

\section{Junction Condition}\label{sec:VII}

The initial research on the Junction condition was carried out by Sen \cite{Sen/1924}. The drawbacks and potential improvements of Den's technique were then discussed by Lanczos \cite{Lanczos/1924}. Despite being a little intricate, Darmois was the first to pinpoint the exact junction condition \cite{Darmois/1927}. For the sake of this work, we follow the simpler and more fundamentally based advice made by Israel \cite{Israel/1966, Israel/1967}.

Since, as was previously noted, the condition along the boundary must match two separate metrics across the narrow shell, we derive the solutions using the Israel junction condition. We use the junction condition in general because the metric along hypersurfaces needs to be differentiable and continuous. As a result, we also verify the results of the boundary condition by looking at the Riemann curvature tensor and the Christoffel symbol because, in the thin shell, we need to use the discontinuity across the junction to find the potential across the thin shell.

Moreover, we note that $\sum$ would represent the three-manifold, or thin shell; $\vartheta^+$ would represent it outside, and $\vartheta^-$ would represent it within. Consequently, $\vartheta^+ \cup \sum \cup \vartheta^-$ would equal the whole space-time. We can also concentrate on the surface energy density $\varsigma$ and surface pressure $p$.

Here are the internal solutions we have.

\begin{equation*}
  ds^2=-e^{A} dt^2+e^{B} dr^2 + r^2(d \theta^2 + \sin^2 \theta d\phi^2).
\end{equation*}

To find the external solutions, we use the format's spherically symmetric solution.

\begin{equation} \label{eq:70}
ds^2= -F(r) dt^2  + F(r)^{-1} dr^2 +r^2(d \theta^2  
+ \sin^2 \theta d\phi^2).
\end{equation}


Here, we observe that the outer metric of a static spherically symmetric gravastar is always the Reissner-Nordstrom metric due to Birkhoff's theorem.
Notably, the space-like component in both scenarios exhibits spherical symmetry, and the FLRW metric may be obtained on the boundary.

\begin{equation} \label{eq:71}
    ds^2=-d\tau^2 + \textbf{a}(t) d\Omega^2.
\end{equation}

Using the formula for the first junction condition now, we get
\begin{equation}\label{eq:72}
     K^{\pm}_{ij}= -n^{\pm}_{\nu}\left(\frac{\partial^2 x^{\nu}}{\partial \phi^{i} \partial \phi^{j}}+ \Gamma^l_{km} \frac{\partial x^l}{\partial \phi^i} \frac{\partial x^m}{\partial \phi^j} \right), 
\end{equation}

where the intrinsic coordinate in the shell region is denoted by $\phi$, and the two-sided unit normal to the surface is represented by $n^{\pm}$.
\begin{equation}\label{eq:73}
    n^{\pm}=\pm \left|g^{lm} \frac{\partial f}{\partial x^{l}} \frac{\partial f}{\partial x^{m}} \right|^{-1/2} \frac{\partial f}{\partial x^{\nu}}, 
\end{equation}
where $n$ is a unit time-like 4-vector that satisfies the equation $n^{\gamma} n_{\gamma} = 1$.
We must use the Lanczos equation to determine the surface tension and pressure for the thin shell to stay stable.
\begin{equation}\label{eq:74}
    S_{ij}=-\frac{1}{8 \pi}(k_{ij}-\delta_{ij} k_{\gamma \gamma}) .
\end{equation}

Here $i, j =0,2,3$. Since $r$ is constant at the shell, $S_{ij}=diag(-\varsigma, p)$ expresses the surface energy tensor. Thus, the following equations can be used to determine the surface energy density $\varsigma$ and pressure $p$ at the junction surface $r=\textbf{a}$:

\begin{equation} \label{eq:75}
     \varsigma=-\frac{1}{4 \pi \textbf{a} }\left[\sqrt {f}\right]^{+}_{-}, 
\end{equation}
and
\begin{equation}\label{eq:76}
         p=-\frac{\varsigma}{2}+\frac{1}{16 \pi}\left[\frac{f^{\prime}}{\sqrt {f}}\right]^{+}_{-} .
\end{equation}



The aforementioned formulas can be used to derive the expressions for the previously stated quantities.
By observing that the energy-momentum has a conservation relation, we may compute the potential $V(r)$ as \cite{M. Sharif, S. D. Forghani}

\begin{equation} \label{eq:77}
    \frac{d}{d\tau}(\varsigma \phi)=p\frac{d\phi}{d\tau}=0\,,
\end{equation}

where $\phi = 4\pi \textbf{a}^2$. Using the conservation equation above, one can determine 
\begin{equation}\label{eq:78}
    \varsigma '=-\frac{2}{\textbf{a}}(\varsigma+p)\,.
\end{equation}

Observing that the final equation takes the form $\dot{\textbf{a}}^2+V(\textbf{a})$, we may obtain, using the prescription from \cite{eric}.

\begin{equation} \label{eq:79}
    V(\textbf{a})=\frac{f(\textbf{a})}{2}+\frac{F(\textbf{a})}{2}-\frac{(f(\textbf{a})-F(\textbf{a}))^2}{64\textbf{a}^2\pi^2\varsigma^2}-4\textbf{a}^2\pi^2\varsigma^2. 
\end{equation}
\subsubsection{Reissner-Nordstrom metric}
\begin{equation} \label{eq:80}
   \varsigma=-\frac{\sqrt{\frac{Q^2}{\textbf{a}^2}-\frac{2 M}{\textbf{a}}+1}-\sqrt{-\left(-\frac{2 M}{R}+\frac{Q^2}{R^2}+1\right)^{\frac{\textbf{a}^2}{R^2}}-c_1}}{4 \pi  \textbf{a}},
\end{equation}

\begin{multline} \label{eq:81}
    p=\frac{\sqrt{\frac{Q^2}{\textbf{a}^2}-\frac{2 M}{\textbf{a}}+1}-\sqrt{-\left(-\frac{2 M}{R}+\frac{Q^2}{R^2}+1\right)^{\frac{\textbf{a}^2}{R^2}}-c_1}}{8 \pi  \textbf{a}}\\ +\frac{\frac{2 \textbf{a} \left(-\frac{2 M}{R}+\frac{Q^2}{R^2}+1\right)^{\frac{\textbf{a}^2}{R^2}} \log \left(-\frac{2 M}{R}+\frac{Q^2}{R^2}+1\right)}{R^2 \left(-\left(-\frac{2 M}{R}+\frac{Q^2}{R^2}+1\right)^{\frac{\textbf{a}^2}{R^2}}-c_1\right)}+\frac{\frac{2 M}{\textbf{a}^2}-\frac{2 Q^2}{\textbf{a}^3}}{\sqrt{\frac{Q^2}{\textbf{a}^2}-\frac{2 M}{\textbf{a}}+1}}}{16 \pi }.
\end{multline}

\subsubsection{Bardeen black holes}

\begin{multline}\label{eq:82}
    \varsigma=\frac{-1}{4 \pi  \textbf{a}}\left(\sqrt{-\frac{2 \textbf{a}^2 M}{\left(\textbf{a}^2+e^2\right)^{3/2}}+\frac{Q^2}{\textbf{a}^2}+1} \right.\\ \left.
    -\sqrt{-\left(-\frac{2 M R^2}{\left(e^2+R^2\right)^{3/2}} +\frac{Q^2}{R^2}+1\right)^{\frac{\textbf{a}^2}{R^2}}-c_1}\right),
\end{multline}

\begin{widetext}
\begin{multline}\label{eq:83}
    p=\frac{\sqrt{-\frac{2 \textbf{a}^2 M}{\left(\textbf{a}^2+e^2\right)^{3/2}}+\frac{Q^2}{\textbf{a}^2}+1}-\sqrt{-\left(-\frac{2 M R^2}{\left(e^2+R^2\right)^{3/2}}+\frac{Q^2}{R^2}+1\right)^{\frac{\textbf{a}^2}{R^2}}-c_1}}{8 \pi  \textbf{a}} \\
    +\frac{\frac{2 \textbf{a} \left(-\frac{2 M R^2}{\left(e^2+R^2\right)^{3/2}}+\frac{Q^2}{R^2}+1\right)^{\frac{\textbf{a}^2}{R^2}} \log \left(-\frac{2 M R^2}{\left(e^2+R^2\right)^{3/2}}+\frac{Q^2}{R^2}+1\right)}{R^2 \left(-\left(-\frac{2 M R^2}{\left(e^2+R^2\right)^{3/2}}+\frac{Q^2}{R^2}+1\right)^{\frac{\textbf{a}^2}{R^2}}-c_1\right)}+\frac{-\frac{2 Q^2}{\textbf{a}^3}-\frac{4 \textbf{a} M}{\left(\textbf{a}^2+e^2\right)^{3/2}}+\frac{6 \textbf{a}^3 M}{\left(\textbf{a}^2+e^2\right)^{5/2}}}{\sqrt{-\frac{2 \textbf{a}^2 M}{\left(\textbf{a}^2+e^2\right)^{3/2}}+\frac{Q^2}{\textbf{a}^2}+1}}}{16 \pi }.
\end{multline}
\end{widetext}

\subsubsection{Hayward black holes}

\begin{equation} \label{eq:84}
    \varsigma=\frac{-1}{4 \pi  \textbf{a}}\left( \sqrt{\frac{Q^2}{\textbf{a}^2}-\frac{2 \textbf{a}^2 M}{\textbf{a}^3+2 l^2 M}+1} 
    -\sqrt{-\left(-\frac{2 M R^2}{2 l^2 M+R^3}+\frac{Q^2}{R^2}+1\right)^{\frac{\textbf{a}^2}{R^2}}-c_1} \right),
\end{equation}

\begin{widetext}
\begin{multline}\label{eq:85}
    p=\frac{\sqrt{\frac{Q^2}{\textbf{a}^2}-\frac{2 \textbf{a}^2 M}{\textbf{a}^3+2 l^2 M}+1}-\sqrt{-\left(-\frac{2 M R^2}{2 l^2 M+R^3}+\frac{Q^2}{R^2}+1\right)^{\frac{\textbf{a}^2}{R^2}}-c_1}}{8 \pi  \textbf{a}}\\
    +\frac{\frac{2 l \left(-\frac{2 M R^2}{2 \textbf{a}^2 M+R^3}+\frac{Q^2}{R^2}+1\right)^{\frac{l^2}{R^2}} \log \left(-\frac{2 M R^2}{2 \textbf{a}^2 M+R^3}+\frac{Q^2}{R^2}+1\right)}{R^2 \left(-\left(-\frac{2 M R^2}{2 \textbf{a}^2 M+R^3}+\frac{Q^2}{R^2}+1\right)^{\frac{l^2}{R^2}}-c_1\right)}+\frac{-\frac{4 \textbf{a} M}{\textbf{a}^3+2 l^2 M}-\frac{2 Q^2}{\textbf{a}^3}+\frac{6 \textbf{a}^4 M}{\left(\textbf{a}^3+2 l^2 M\right)^2}}{\sqrt{-\frac{2 l^2 M}{2 \textbf{a}^2 M+l^3}+\frac{Q^2}{l^2}+1}}}{16 \pi }.
\end{multline}
\end{widetext}

\begin{figure}[H]
    \centering
    \includegraphics[scale=0.45]{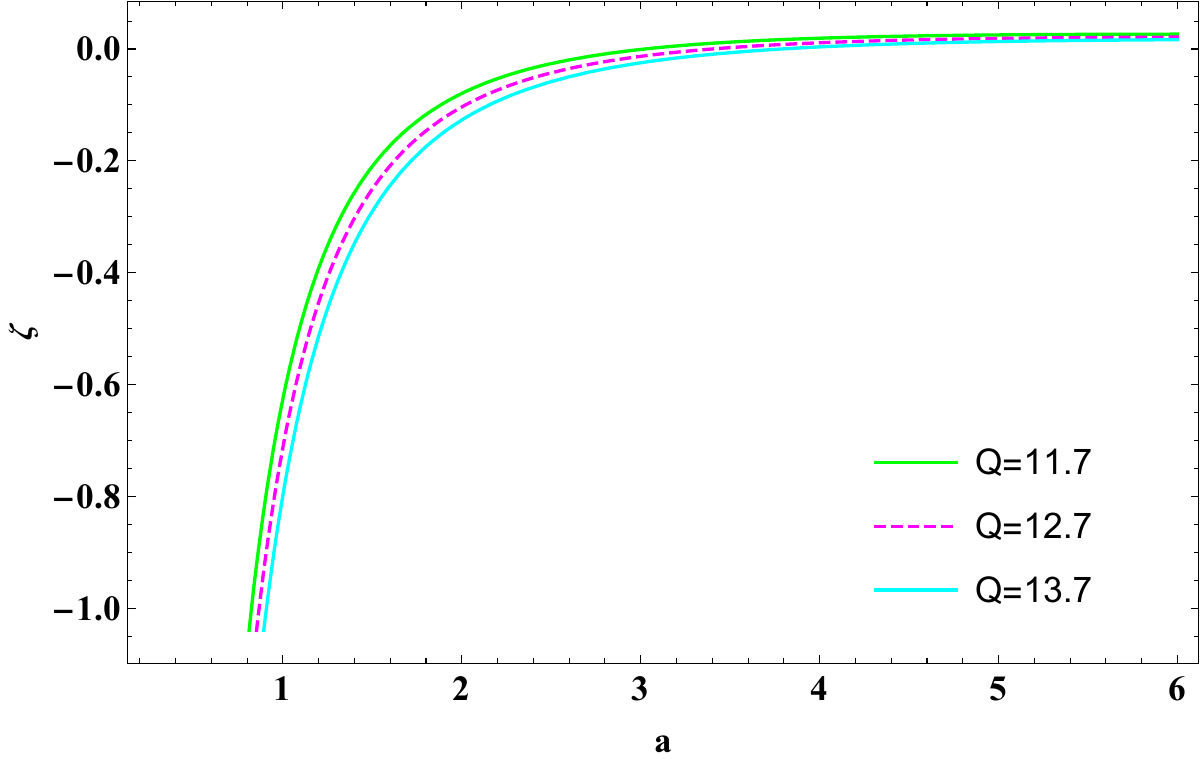}
    \caption{Variation of surface energy density inside the thin shell in R-N metric.}
    \label{fig-1}
\end{figure}

\begin{figure}[H]
    \centering
    \includegraphics[scale=0.45]{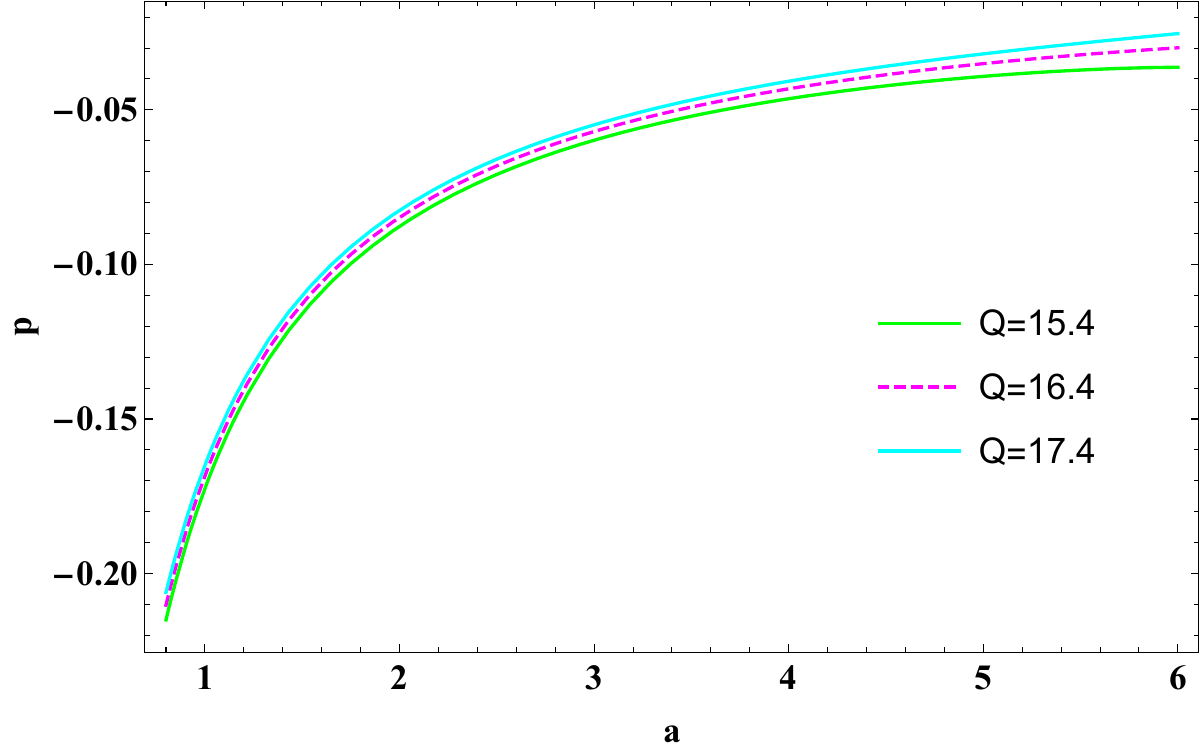}
    \caption{Variation of surface pressure inside the thin shell in R-N metric.}
    \label{fig-2}
\end{figure}

\begin{figure}[H]
    \centering
    \includegraphics[scale=0.45]{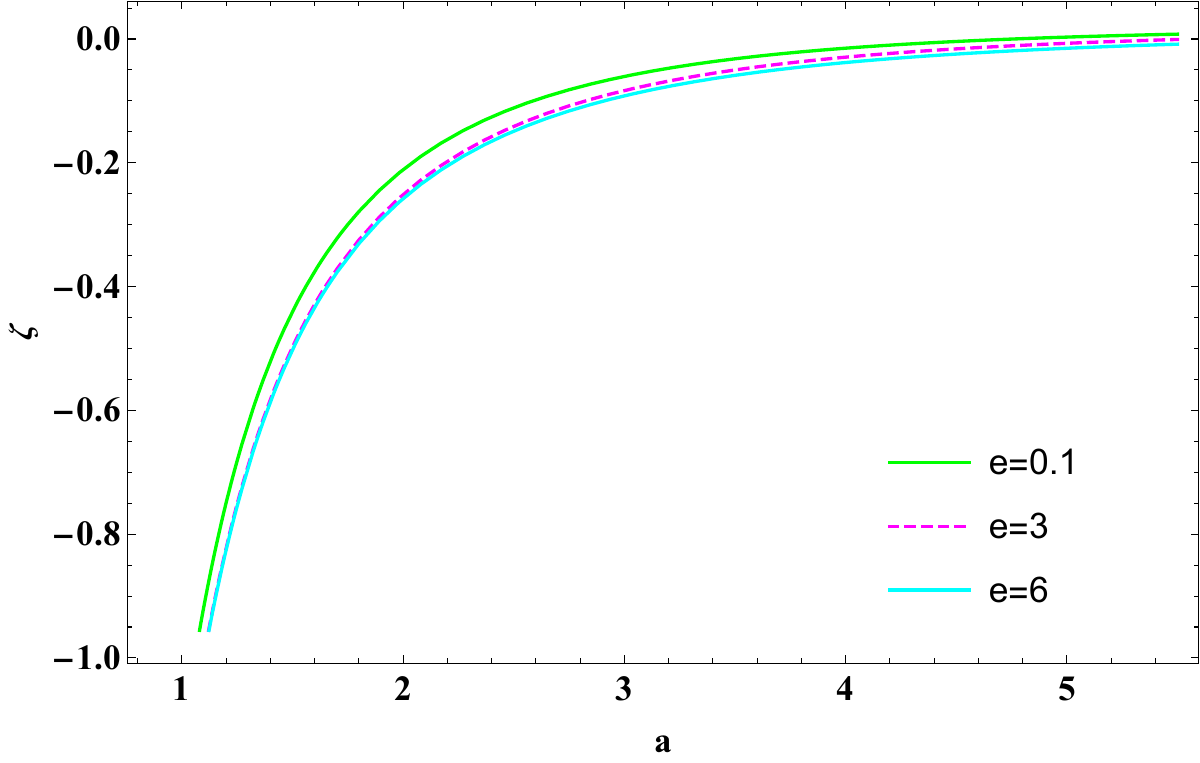}
    \caption{Variation of surface energy density for different values of $e$ inside the thin shell in Bardeen black hole.}
    \label{fig-3}
\end{figure}

\begin{figure}[H]
    \centering
    \includegraphics[scale=0.45] {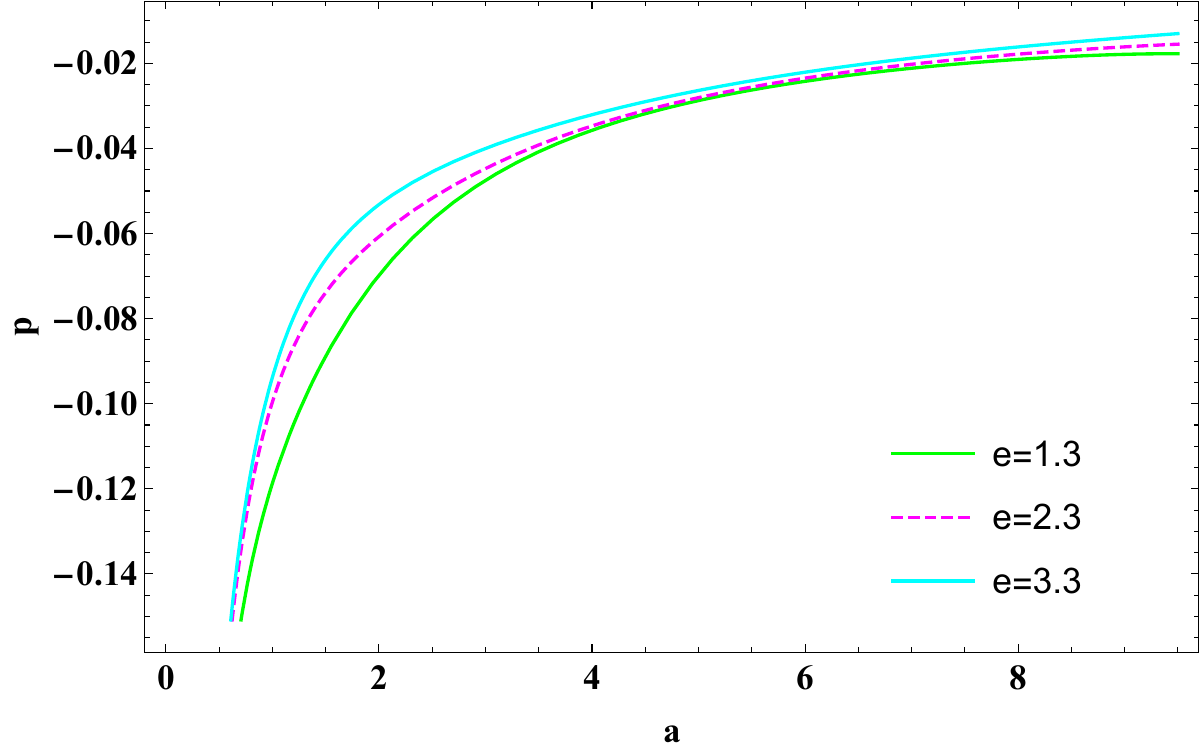}
    \caption{Variation of surface pressure for different values of $e$ inside the thin shell in Bardeen black hole.}
    \label{fig-4}
\end{figure}

\begin{figure}[H]
    \centering
    \includegraphics[scale=0.45]{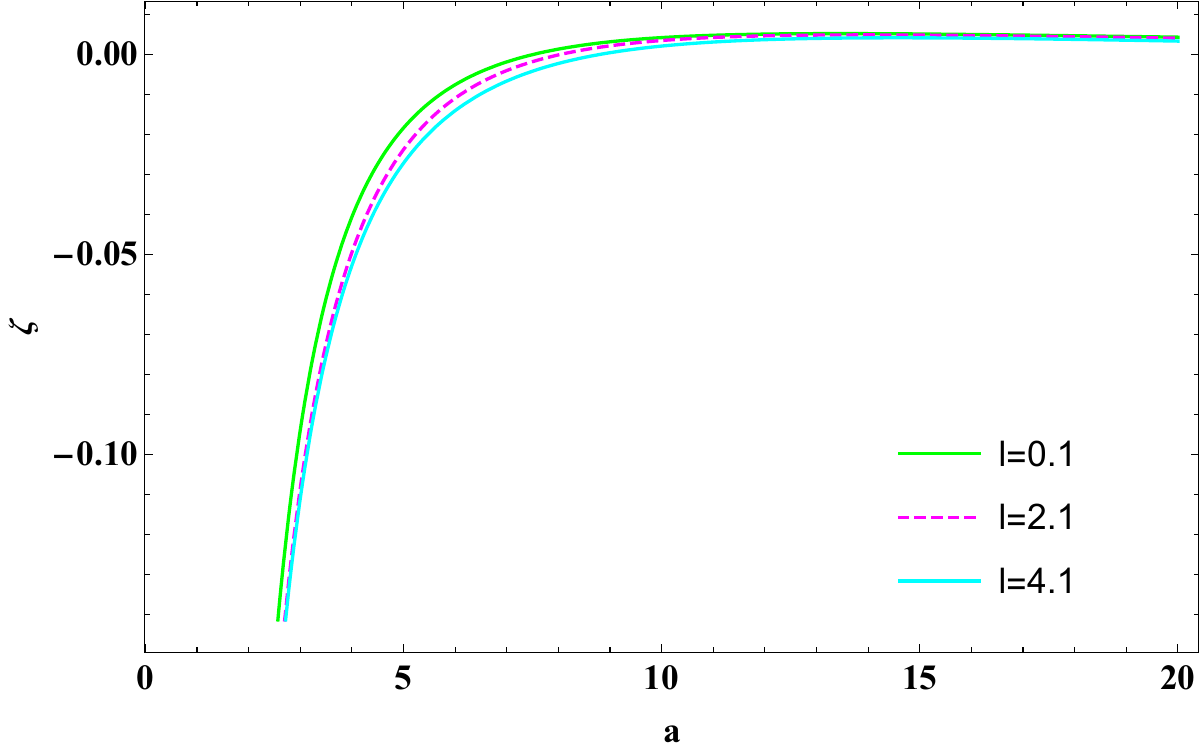}
    \caption{Variation of surface energy density for different values of $l$ inside the thin shell in Hayward black hole.}
    \label{fig-5}
\end{figure}

\begin{figure}[H]
    \centering
    \includegraphics[scale=0.45]{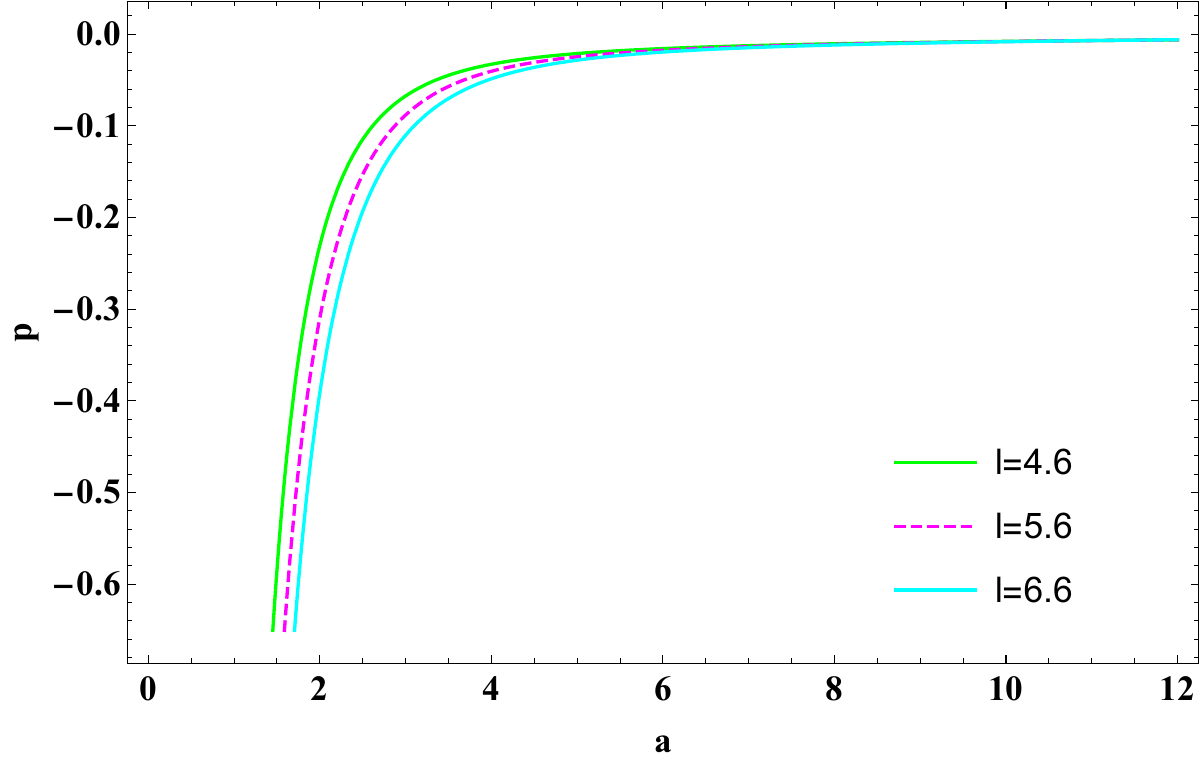}
    \caption{Variation of surface pressure for different values of $l$ inside the thin shell in Hayward black hole.}
    \label{fig-6}
\end{figure}

\begin{figure}[H]
    \centering
    \includegraphics[scale=0.45]{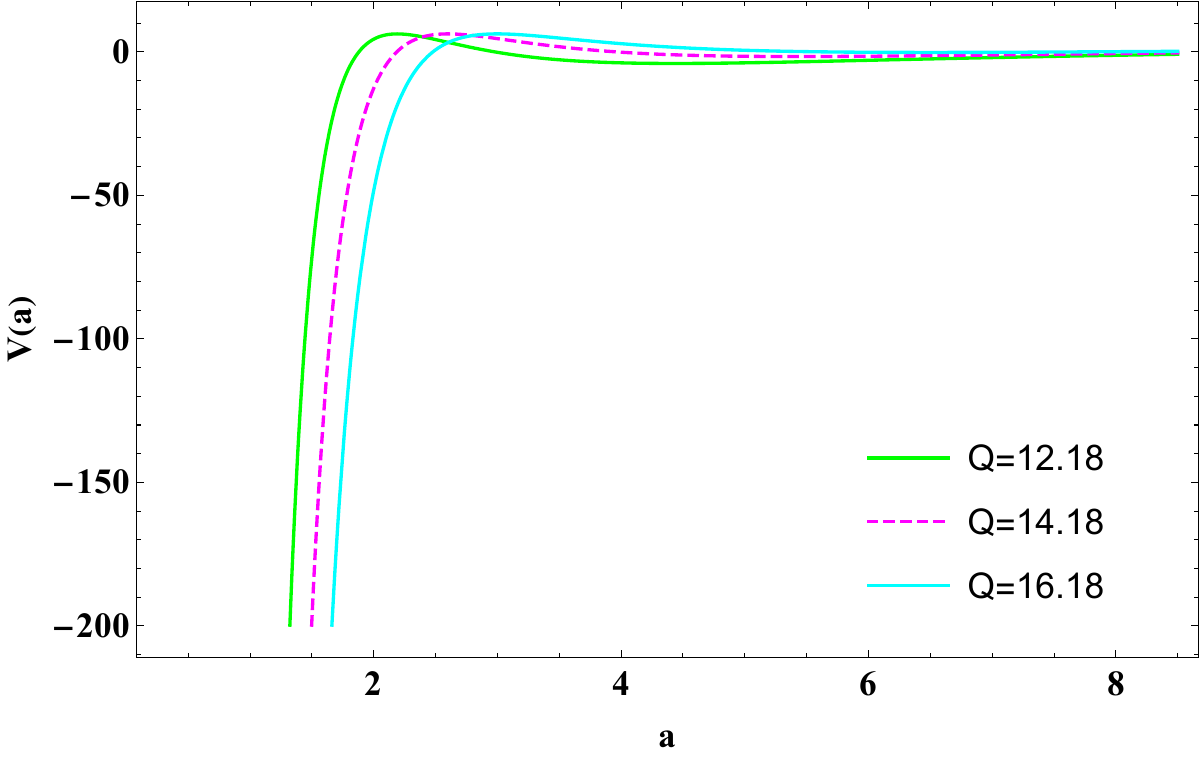}
    \caption{Potential for shell outside R-N metric.}
    \label{fig-7}
\end{figure}

\begin{figure}[H]
    \centering
    \includegraphics[scale=0.45]{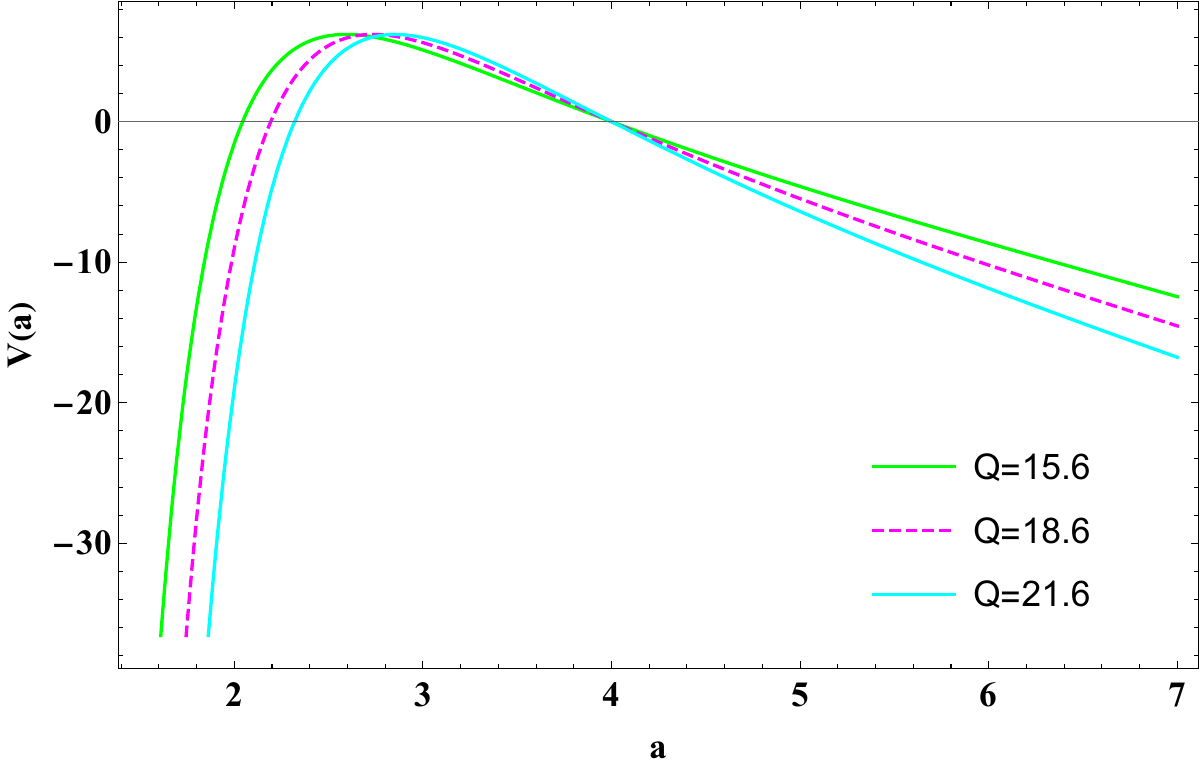}
    \caption{Potential for shell outside Bardeen black hole.}
    \label{fig-8}
\end{figure}

\begin{figure}[H]
    \centering
    \includegraphics[scale=0.45]{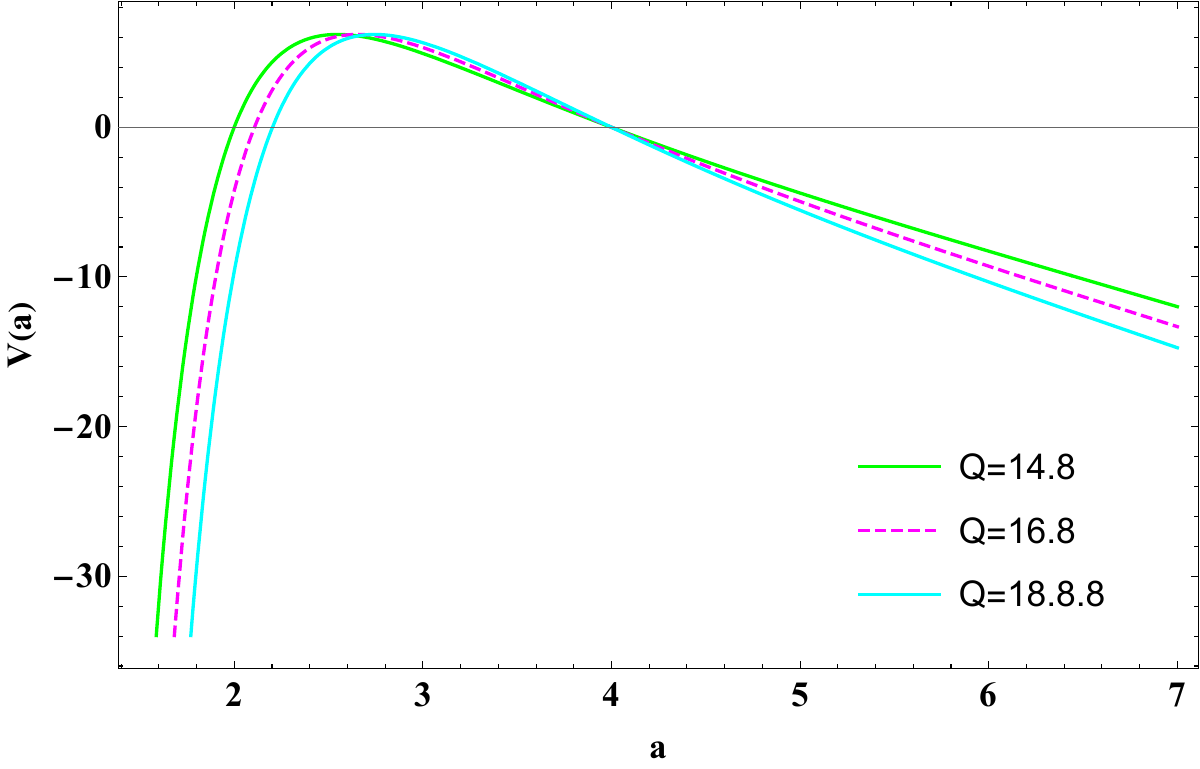}
    \caption{Potential for shell outside Hayward black hole.}
    \label{fig-9}
\end{figure}

\section{PHYSICAL FEATURES OF THE MODEL}\label{sec:VIII}
The equation of state, the proper length, the entropy, stability, and the energy levels inside the shell's region are some of the physical properties of the constructed model, and they will be covered in this section.  Because the constructed geometry of the gravastar matches two distinct space-times, the stiff matter fluid moves along these space-times through the gravastar's shell region. It will also study the effects of the electromagnetic field on various physical properties of the charged gravastar in the $f(\mathcal{Q})$ gravity.

\subsection{Proper length}
The shell lies between the point where two space-times converge, as per the theories of Mazur and Mottola \cite{Mottola/2002, Mazur/2004}. The shell's length varies from the phase border $r_2=d+\epsilon$ between the outside space-time and the intermediate thin shell to the phase barrier $r_1=d$ between the inner area and the intermediate thin shell.  Between the interface boundaries, the appropriate thickness of the shell is 
    \begin{multline} \label{eq:86}
        \begin{gathered}
           \textit{l}=\int^{d+\epsilon}_{d} \sqrt{e^{B(r)}} dr \\ = \int^{d+\epsilon}_{d}\frac{1}{\sqrt{-e^{-X r^2} -c_1}} dr \\=\int_d^{d+\epsilon } \frac{1}{\sqrt{-c_1-\left(-\frac{2 M}{R}+\frac{Q^2}{R^2}+1\right)^{\frac{r^2}{R^2}}}} \, dr\\=\int^{d+\epsilon}_{d} \frac{1}{f(r)} dr.
        \end{gathered}
    \end{multline}

Where $f(r)=\sqrt{-c_1-\left(-\frac{2 M}{R}+\frac{Q^2}{R^2}+1\right)^{\frac{r^2}{R^2}}}$

Currently, evaluating the integral given in equation \eqref{eq:86} is somewhat difficult. Thus, let us choose $\frac{d f(r)}{dr}=\frac{1}{f(r)}$ to solve the preceding integral. Thus, we obtain,

\begin{equation} \label{eq:87}
    \textit{l}= f(d+\epsilon)-f(d). 
\end{equation}

After extending $f(d+\epsilon)$ in the Taylor series around `$d$', we obtain from equation \eqref{eq:87}, maintaining the linear order of $\epsilon$.

\begin{equation}\label{eq:88}
    \textit{l}= \epsilon \frac{d f(r)}{dr} \approx  \frac{\epsilon}{\sqrt{-c_1-\left(-\frac{2 M}{R}+\frac{Q^2}{R^2}+1\right)^{\frac{r^2}{R^2}}}}. 
\end{equation}

\begin{figure}[h]
    \centering
    \includegraphics[scale=0.45]{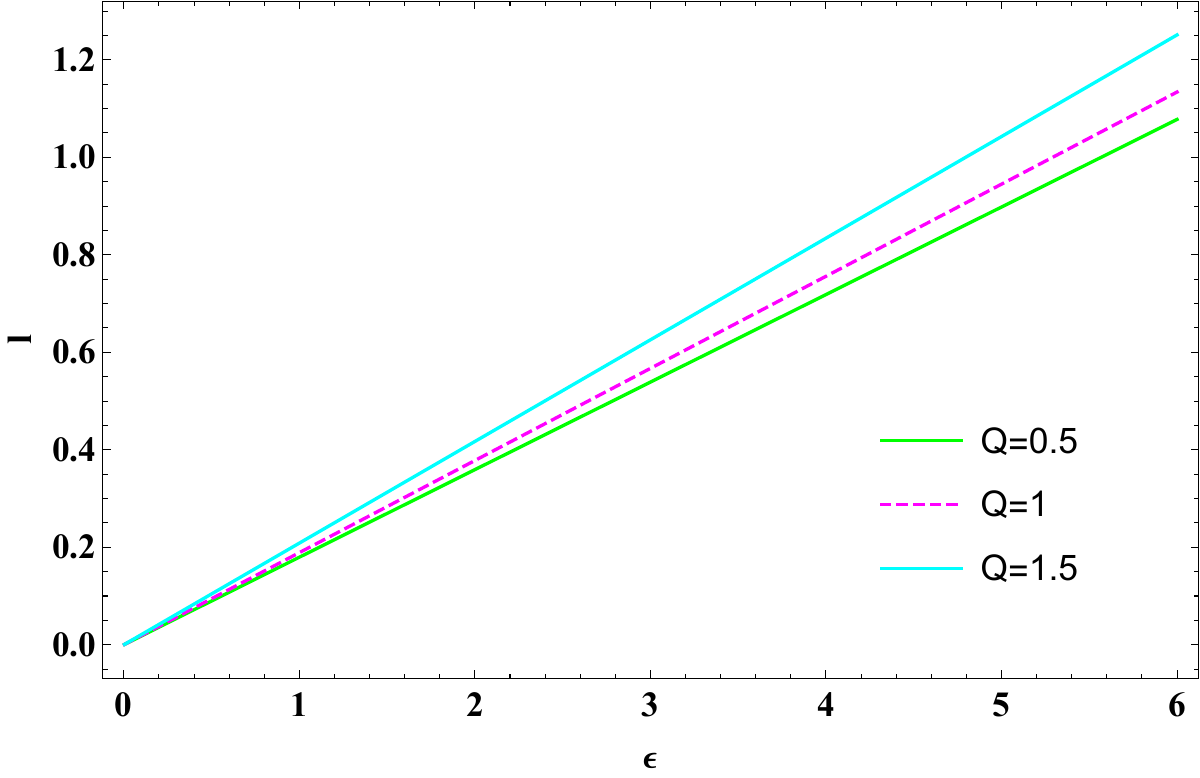}
    \caption{Variation of proper length inside the thin shell}
    \label{fig-10}
\end{figure}

Because of $\epsilon$'s extremely small value, the higher-order terms of the exponential can be ignored. The proper length variation for the thin shell radius is shown in Fig.\eqref{fig-10}.

\subsection{Entropy}
The stable configuration for a single condensate area is found in the innermost region of the gravastar, which has zero entropy density. The formula can be utilized to calculate entropy on the intermediate thin shell, as per the findings of Mazur and Mottola \cite{Mottola/2002,Mazur/2004}.

\begin{equation} \label{eq:89}
   \mathcal{S}= \int^{d+\epsilon}_{d} 4 \pi r^2 s(r) \sqrt{e^{B(r)}} dr.
\end{equation}

The equation of state, $p = \rho = \frac{k^2}{8 \pi G}$, gives the entropy of the thin shell. For dimensional reasons, $G$ has been included; therefore, ${k}^2$ is a dimensionless constant. The local specific entropy density of a relativistic fluid with zero chemical potential is $sT = p + \rho$, according to the standard thermodynamic Gibbs relation.
\begin{equation} \label{eq:90}
    s(r)= \frac{k^2 {K_B}^2 T(r)}{4 \pi \hbar^2 G}=\frac{k K_B}{\hbar}\left(\frac{p}{2 \pi G}\right)^{\frac{1}{2}}.
\end{equation}

Geometrical units $G=1$ have been considered in this work. Now equation \eqref{eq:89} becomes

\begin{multline}\label{eq:91} 
S=\int^{d+\epsilon}_{d} 4 \pi r^2 k \left( \frac{K_B}{\hbar}\right) \left(\frac{p}{2 \pi}\right)^{\frac{1}{2}} \sqrt{e^{B(r)}} dr \\ =\int^{d+\epsilon}_{d} 2 \sqrt{2 \pi} r^2 k \left( \frac{K_B} {\hbar}\right) \sqrt{p e^{B(r)}} dr  .      
\end{multline}

The above equation can be written as
\begin{equation} \label{eq:92}
    S= 2 \sqrt{2 \pi}r^2 k \left(  \frac{K_B}{\hbar}\right) N,
\end{equation}
where 
\begin{equation} \label{eq:93}
    N= \int^{d+\epsilon}_{d} D(r) dr,
\end{equation}

\begin{widetext}
\begin{equation}\label{eq:94}
    D(r)=\frac{\sqrt{\frac{\frac{\left(c_1+\left(\frac{-2 M R+Q^2+R^2}{R^2}\right)^{\frac{r^2}{R^2}}\right) \left(r \left(\sqrt{\frac{-2 M r+Q^2+r^2}{r^2}} \sqrt{-c_1-\left(\frac{-2 M R+Q^2+R^2}{R^2}\right)^{\frac{r^2}{R^2}}}-1\right)+M\right)}{\sqrt{\frac{-2 M r+Q^2+r^2}{r^2}}}+\frac{r^3 \left(\frac{-2 M R+Q^2+R^2}{R^2}\right)^{\frac{r^2}{R^2}} \log \left(\frac{-2 M R+Q^2+R^2}{R^2}\right)}{R^2}}{r^2 \left(c_1+\left(\frac{-2 M R+Q^2+R^2}{R^2}\right)^{\frac{r^2}{R^2}}\right)^2}}}{2 \sqrt{2 \pi }}.
\end{equation}
\end{widetext}

According to equation \eqref{eq:93}, evaluating the integral is currently quite challenging. Assuming $F(r)$ to be the primitive of $D(r)$, we can compute the integral mentioned above. Following that, equation \eqref{eq:93} becomes: after using the integral calculus fundamental theorem.

\begin{equation}\label{eq:95}
    N=[F(r)]^{d+\epsilon}_{d}=F(d+\epsilon)-F(d).
\end{equation}

After retaining the linear order of $\epsilon$ from equation \eqref{eq:95} and expanding $F(d+\epsilon)$ in the Taylor series around `$d$', we find, from \eqref{eq:92}, that
\begin{widetext}
\begin{equation}\label{eq:96}
   S= \frac{k K_B r^2 \epsilon  \sqrt{\frac{\frac{\left(c_1+\left(\frac{-2 M R+Q^2+R^2}{R^2}\right)^{\frac{r^2}{R^2}}\right) \left(r \left(\sqrt{\frac{-2 M r+Q^2+r^2}{r^2}} \sqrt{-c_1-\left(\frac{-2 M R+Q^2+R^2}{R^2}\right)^{\frac{r^2}{R^2}}}-1\right)+M\right)}{\sqrt{\frac{-2 M r+Q^2+r^2}{r^2}}}+\frac{r^3 \left(\frac{-2 M R+Q^2+R^2}{R^2}\right)^{\frac{r^2}{R^2}} \log \left(\frac{-2 M R+Q^2+R^2}{R^2}\right)}{R^2}}{r^2 \left(c_1+\left(\frac{-2 M R+Q^2+R^2}{R^2}\right)^{\frac{r^2}{R^2}}\right)^2}}}{h}.
\end{equation}
\end{widetext}

\begin{figure*}[t]
    \includegraphics[scale=0.41]{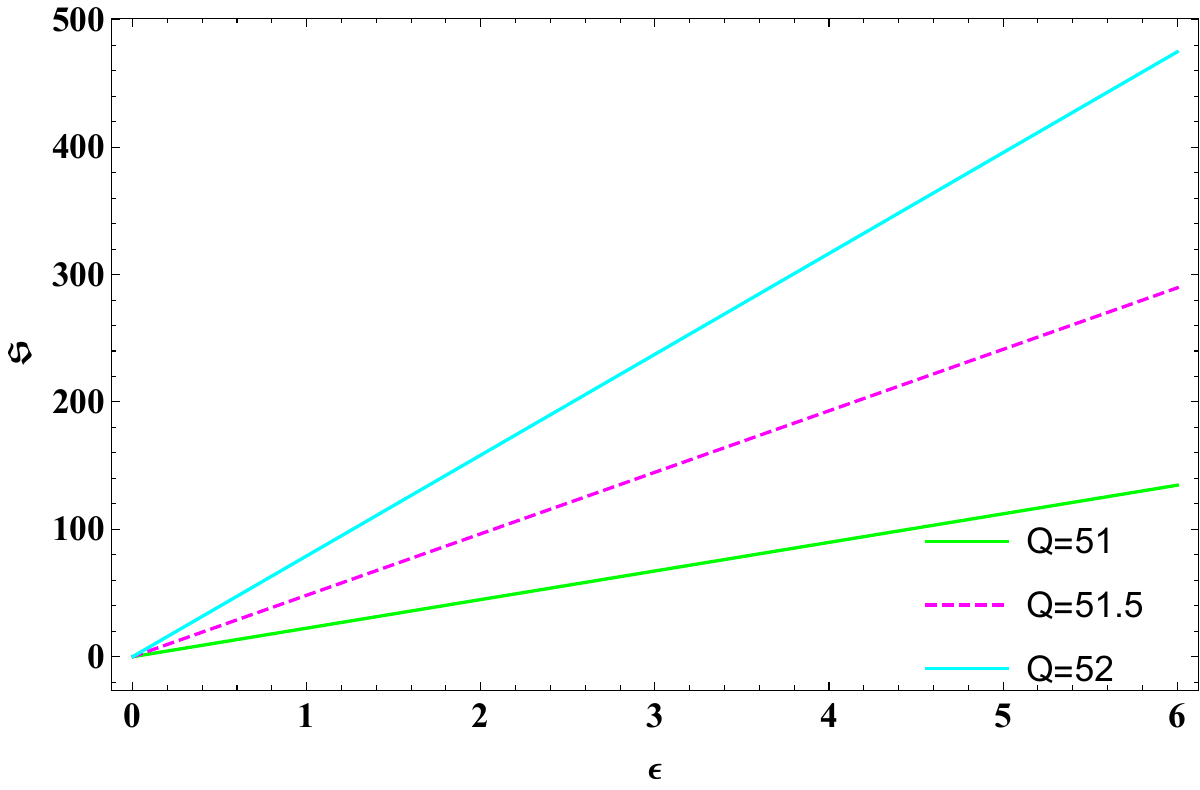}\,\,\,\,\,\,\,\,\,\,\,\,
    \includegraphics[scale=0.41]{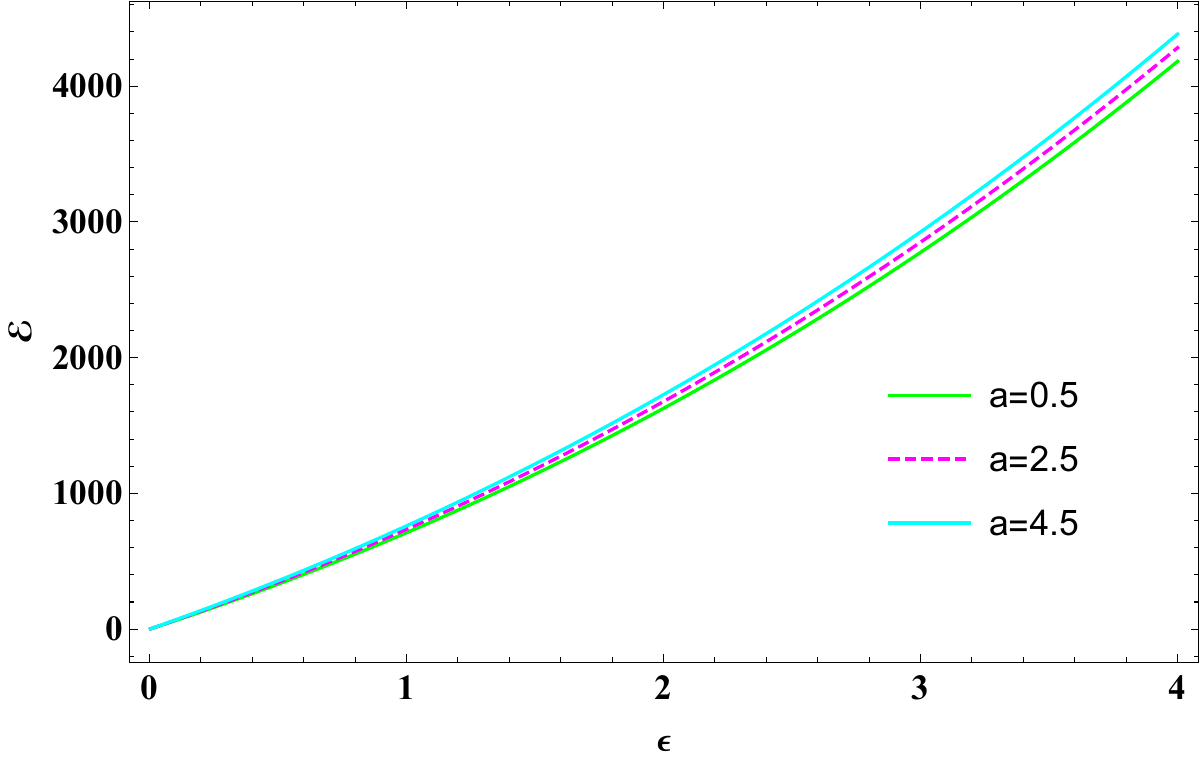}
    \caption{Variation of entropy with respect to thickness (left plot) and variation of energy inside the thin shell (right plot)}
    \label{fig-11}
\end{figure*}

Thus, the entropy expression for our proposed model was successfully obtained. The entropy change with respect to the thin shell radius is shown in the left plot of Fig.\eqref{fig-11}.

\subsection{Energy}

The shell's energy can be computed using the formula.
\begin{widetext}
\begin{multline}\label{eq:97}
\begin{gathered}
    \mathcal{E} =\int^{d+\epsilon}_{d} 4 \pi r^2 (\rho+ 2 \pi E^2) dr \\
    =\frac{2 \pi ^{3/2} a Y \left(\text{erf}\left(d \sqrt{X}\right)-\text{erf}\left(\sqrt{X} (d+\epsilon )\right)\right)}{X^{3/2}}
    +\frac{4 \pi  a e^{-X (d+\epsilon )^2} \left(d (X-Y) \left(e^{X \epsilon  (2 d+\epsilon )}-1\right)+\epsilon  \left(X \left(e^{X (d+\epsilon )^2}-1\right)+Y\right)\right)}{X} \\  +\frac{2}{3} \pi  b \epsilon  \left(3 d^2+3 d \epsilon +\epsilon ^2\right).
    \end{gathered}
\end{multline}
\end{widetext}
 
This shows a clear correlation between the energy and shell thickness. For various values of $a$, the nature of energy inside the thin shell is seen in the right plot of Fig. \eqref{fig-11}.


\subsection{The EoS parameter}
The equation of state's state parameter, $\omega$, can be stated as \cite{Pradhan/2023} 
\begin{widetext}
\begin{equation}\label{eq:98}
    \omega=\frac{p}{\varsigma}=\frac{1}{4} \left(-\frac{\textbf{a} \left(\frac{2 \left(\textbf{a} M-Q^2\right)}{\textbf{a}^3 \sqrt{\frac{\textbf{a}^2-2 \textbf{a} M+Q^2}{\textbf{a}^2}}}-\frac{2 \textbf{a} \left(\frac{-2 M R+Q^2+R^2}{R^2}\right)^{\frac{\textbf{a}^2}{R^2}} \log \left(\frac{-2 M R+Q^2+R^2}{R^2}\right)}{R^2 \left(\left(\frac{-2 M R+Q^2+R^2}{R^2}\right)^{\frac{\textbf{a}^2}{R^2}}+c_1\right)}\right)}{\sqrt{\frac{\textbf{a}^2-2 \textbf{a} M+Q^2}{\textbf{a}^2}}-\sqrt{-\left(\frac{-2 M R+Q^2+R^2}{R^2}\right)^{\frac{\textbf{a}^2}{R^2}}-c_1}}-2\right).
\end{equation}
\end{widetext}

\begin{figure}[h]
    \includegraphics[scale=0.4]{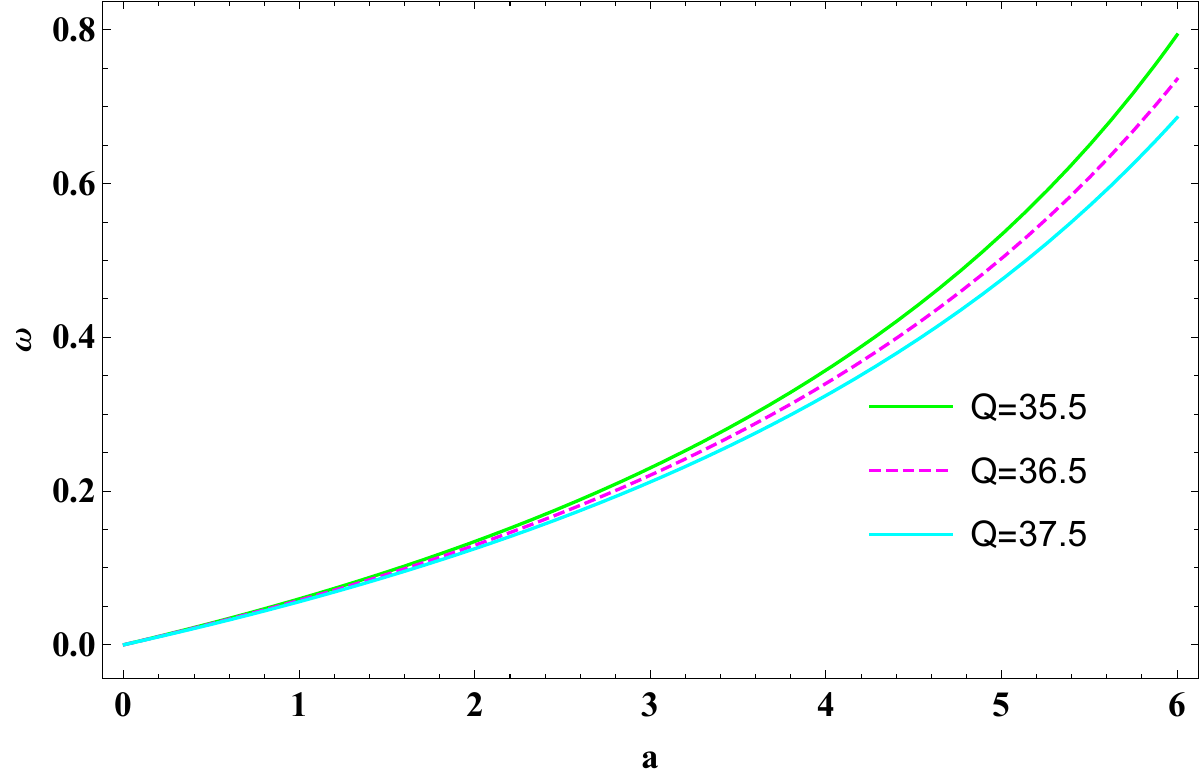}
    \caption{Variation of EoS with respect to \textbf{a}.}
    \label{fig-12}
\end{figure}



\subsection{Stability}
\begin{figure}[H]
    \includegraphics[scale=0.4]{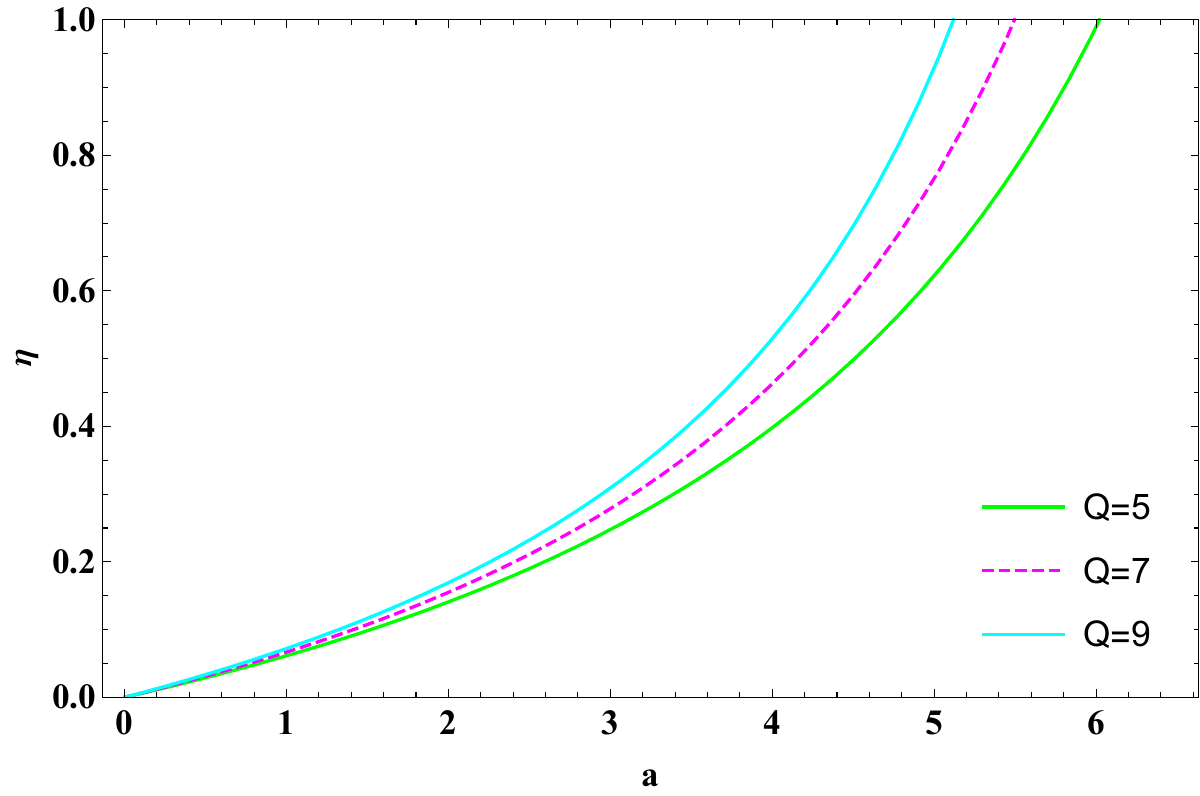}
    \caption{Variation of $\eta$ with respect to \textbf{a} for different values of $Q$.}
    \label{fig-13}
\end{figure}

As we have calculated the pressure ($p$) and energy density ($\varsigma$) for the thin-shell from the equations \eqref{eq:80} and \eqref{eq:81}, We note that for a general fluid, the speed of ``sound" (or the perturbation of wave through that distorted media) can be given as:
\begin{equation}\label{eq:99}
    \eta=\frac{p'}{\varsigma'}
\end{equation}
Here, we can evaluate the expression at $r=\textbf{a}$ to get the speed of sound in the thin-shell. It was first proposed by Poisson and Visser \cite{Poisson/1995} that from causality reasons, it could be expected that the speed of sound should not exceed the speed of light (here $c=1$ in natural unit). So one can expect from the causality condition that the $\eta$ would satisfy $0\leq \eta\leq 1$.\\
As we can see from Fig-\ref{fig-13}, $0\leq \eta\leq 1$ gives a bound on the thickness of the shell ($\textbf{a}$) for various values of $Q$. Here, we have plotted the three different $Q$ (i.e. $Q=5,7,9$), and the plot shows for each case the allowed thickness ($\textbf{a}$) for each value of $Q$.\\
The stability of analysis of gravastar via bound on the speed of sound has been explored in literature quite intensively. For example, Lobo \cite{Lobo/2003} has used this stability tool to give a bound for a wormhole in the presence of a cosmological constant. Other works like \cite{Pramit/2021,Yousaf/2019,Debnath/2021} also study the stability of charged gravastar.\\
However, as mentioned in  \cite{Poisson/1995}, analysing this stability via the speed of sound has its own limitations. For example, near $\omega=1$ (stiff matter region), one can not confidently say that the equation \eqref{eq:99} does actually give the speed of sound. The reason for this is that we do not have a full understanding of the microscopic degrees of freedom of the stiff matter so the speed of sound expression via ordinary fluid argument might not hold.\\
Even though it does not offer sufficient conditions, the stability condition provided a necessary condition for the stability of the thin shell around the gravastar.

\section{Conclusion}\label{sec:IX}
In this study, we've developed a solution for gravastars within the framework of $f(Q)$ gravity, utilizing Krori-Barua metric potential ($ B(r)=Xr^2, A(r)=Yr^2+Z $) as the matter content. Since it is well known that  KB metric produces a singularity-free model. First, we'll briefly explain why gravastars or regular astrophysical objects are considered valuable substitutes for black holes. In our conversations, we've explored the reasoning behind the gravastar theory. Specifically, we examined how, by including quantum effects during the process of gravitational collapse, it's conceivable that remaining scalar fields might undergo a Bose-Einstein condensation. This process could result in the formation of a ``dark star" core, exhibiting characteristics similar to those of dark energy. Additionally, this scenario would produce a surrounding layer of rigid matter (with $\omega=1$), which plays a role in the overall entropy. In this article, we have taken Krori-Barua metric in the interior and shell regions of the gravastar. We've initiated our exploration of gravastar by delving into its three distinct regions: the interior, characterized by an equation of state (EoS) of -1; the shell, which has an EoS of 1; and the exterior region, which is described by the Reissner-Nordstrom metric due to the presence of a non-zero charge. It's also important to highlight that our analysis extends beyond just the Reissner-Nordstrom metrics. We have incorporated external metrics, such as Bardeen and Hayward metrics, into our study.  In this article, we have taken Krori-Barua metric as the core of the gravastar model, and we have taken it in both interior and shell regions, so in section \ref{sec:VI}, we match our internal spacetime to the outer spacetime to finding the value of constants $X, Y$ and Z. In this section, we have obtained a series of relations of the boundary surface $r=R$, based on the continuity of the metric coefficient $g_{tt}, g_{rr}$ and $\frac{\partial g_{tt}}{\partial r}$ in R-N metric, Bardeen black hole and Hayward black hole.
In section \ref{sec:VII},  we applied the Israel junction criteria to the thin shell, ignoring its thickness. Subsequently, we determined the energy density and pressure associated with the thin shell for three distinct metrics. These findings are illustrated in Figures \eqref{fig-1},\eqref{fig-2},\eqref{fig-3},\eqref{fig-4},\eqref{fig-5} and \eqref{fig-6}, respectively.We have also discussed an analysis to determine the potential across the thin shell using the Israel junction conditions.  Our approach involved the application of four distinct types of metrics to explore the practical consequences of our investigation. In Figures \eqref{fig-7}, \eqref{fig-8} and \eqref{fig-9}, we have given a graphical representation of the potential across the shell; it was noted that each exhibited a minimum point, indicating the condition necessary for a circular orbit. This method of expressing the potential has provided us with a mechanism to evaluate the gravastar model by examining the motion of a point-like particle in its vicinity.\\ 
In section \ref{sec:VIII}, to explore the characteristics of gravastars, we adopted the conventional techniques introduced by Mazur \cite{Mazur/2004}, providing a comprehensive insight into gravastar properties. In fig-\eqref{fig-10}, the variation of proper length $(l)$ with respect to thickness $(\epsilon)$ has shown, which shows that it is monotonically increasing as it is expected. In fig-\eqref{fig-11}, the left side plot shows the entropy vs thickness, and the right side plot shows the energy variation inside the thin shell. Fig-\eqref{fig-12} shows the variation of the equation of state parameter with respect to \textbf{a} (from junction condition). Based on Fig-\eqref{fig-13}, we observe that the condition $0 \leq \eta \leq 1$ establishes a limit on the thickness parameter $(\textbf{a})$ across a range of $Q$ values. This figure includes plots for three specific $Q$ values: $5, 7,$ and 9. Each plot illustrates the permissible thickness $(\textbf{a})$ corresponding to each $Q$ value. We hope that our phenomenological study, like the stability vs thickness relation, will be tested via future radio telescopes. 
\\
In the future, one might be interested in exploring more general forms of $f(\mathcal{Q})$ gravity to study these types of gravastar solutions. We would also like to note that there are other forms of modified gravity like $f(R) $ \cite{Amit/2017},$f(G)$ \cite{Shah/2023}, and $f(T)$ \cite{Das/2020}, which have been studied extensively in both the cosmological and the astrophysical contexts. It would be of interest whether these modified gravities give the same thickness vs stability relations as of $f(\mathcal{Q})$ gravity or they differ significantly; in case they do, it would be worth finding out the reason they differ. \\

\section*{\NoCaseChange{Data Availability Statement}}
There are no new data associated with this article.

\section*{\NoCaseChange{Acknowledgments}}
 DM expresses gratitude to the BITS-Pilani, Hyderabad campus, India, for the financial support. PKS  acknowledges the National Board for Higher Mathematics (NBHM) under the Department of Atomic Energy (DAE), Govt. of India for financial support to carry out the Research project No.: 02011/3/2022 NBHM(R.P.)/R \& D II/2152 Dt.14.02.2022 and IUCAA, Pune, India for providing support through the visiting Associateship program.

\end{document}